\documentclass[journal]{IEEEtran}
\usepackage{cite}
\usepackage{amsmath,graphicx}
\usepackage{epstopdf}
\usepackage{xcolor}
\usepackage{amssymb}
\usepackage{pifont}
\usepackage{array}
\usepackage{booktabs}
\usepackage{bm}
\usepackage{multirow}
\usepackage{graphicx}
\usepackage{color}
\usepackage{float}
\usepackage{stfloats}
\usepackage[misc]{ifsym}
\usepackage{stmaryrd}
\usepackage{grffile}
\usepackage{subfig}
\usepackage{makecell}
\usepackage{threeparttable}
\usepackage{enumitem}
\usepackage{enumerate}

\begin{document}

\title{Light Field Image Super-Resolution Using Deformable Convolution}

\author{Yingqian Wang, Jungang Yang, Longguang Wang, Xinyi Ying, Tianhao Wu, Wei An, and Yulan Guo

\thanks{This work was partially supported in part by the National Natural Science Foundation of China (Nos. 61972435, 61401474, 61921001).}

\thanks{Y.~Wang, J.~Yang, L.~Wang, X. Ying, T.~Wu, W.~An, and Y.~Guo are with the College of Electronic Science and Technology, National University of Defense Technology, P. R. China. Y. Guo is also with the School of Electronics and Communication Engineering, Sun Yat-sen University, P. R. China. Emails: \{wangyingqian16, yangjungang, wanglongguang15, yingxinyi18, wutianhao16, anwei, yulan.guo\}@nudt.edu.cn. (Corresponding author: Jungang~Yang, Yulan~Guo)}}

\markboth{IEEE Transactions on Image Processing}%
{Shell \MakeLowercase{\textit{et al.}}: Bare Demo of IEEEtran.cls for IEEE Journals}

\maketitle

\begin{abstract}
  Light field (LF) cameras can record scenes from multiple perspectives, and thus introduce beneficial angular information for image super-resolution (SR).  However, it is challenging to incorporate angular information due to disparities among LF images. In this paper, we propose a deformable convolution network (i.e., LF-DFnet) to handle the disparity problem for LF image SR. Specifically, we design an angular deformable alignment module (ADAM) for feature-level alignment. Based on ADAM, we further propose a collect-and-distribute approach to perform bidirectional alignment between the center-view feature and each side-view feature. Using our approach, angular information can be well incorporated and encoded into features of each view, which benefits the SR reconstruction of all LF images. Moreover, we develop a baseline-adjustable LF dataset to evaluate SR performance under different disparity variations. Experiments on both public and our self-developed datasets have demonstrated the superiority of our method. Our LF-DFnet can generate high-resolution images with more faithful details and achieve state-of-the-art reconstruction accuracy. Besides, our LF-DFnet is more robust to disparity variations, which has not been well addressed in literature.
\end{abstract}


\section{Introduction}\label{introduction}

  \IEEEPARstart{A}{lthough} light field (LF) cameras enable many attractive functions such as post-capture image editing \cite{wang2018selective,xiao2017aliasing,zhang2016plenopatch}, depth sensing \cite{mishiba2020fast,chuchvara2019fast,zhou2019unsupervised,liu2019binocular,shi2019framework,sheng2017geometric}, saliency detection \cite{piao2019saliency,zhang2020light,zhang2019memory,wang2019LFSD,li2014saliency}, and de-occlusion \cite{li2018robust,INRIA,DeOccNet}, the resolution of a sub-aperture image (SAI) is much lower than that of the total sensors. The low spatial resolution problem hinders the development of LF imaging \cite{wu2017overview}. Since high-resolution (HR) images are required in various LF applications, it is necessary to reconstruct HR images from low-resolution (LR) observations, namely, to perform LF image super-resolution (SR).

  To achieve high SR performance, information both within a single view (i.e., spatial information) and among different views (i.e., angular information) is important. Several models have been proposed in early LF image SR methods, such as variational model \cite{wanner2013variational}, Gaussian mixture model \cite{mitra2012light}, and PCA analysis model \cite{farrugia2017super}.  Although different delicately handcrafted image priors have been investigated in these traditional methods \cite{wanner2013variational,mitra2012light,farrugia2017super,GB,ghassab2019light,LFBM5D}, their performance is relatively limited due to their inferiority in spatial information exploitation. In contrast, recent deep learning-based methods \cite{LFCNN2015,LFCNN2017,LF-DCNN,resLF,LFNet,LFSSR,LF-InterNet} enhance spatial information exploitation via cascaded convolutions, and thus achieve improved performance as compared to traditional methods. Yoon et al. \cite{LFCNN2015,LFCNN2017} proposed the first CNN-based method LFCNN for LF image SR. Specifically, sub-aperture images (SAIs) are first super-resolved using SRCNN \cite{SRCNN2015}, and then fine-tuned in pairs to incorporate angular information. Similarly, Yuan et al. \cite{LF-DCNN} super-resolved each SAI separately using EDSR \cite{EDSR}, and then proposed an EPI-enhancement network to refine the results. Although several recent deep learning-based methods \cite{LFNet,LFSSR,resLF,LF-InterNet} have been proposed to achieve the state-of-the-art performance,    the \textit{disparity} issue in LF image SR is still under-investigated.


  In real-world scenes, objects at different depths have different disparity values in LF images. Existing CNN-based LF image SR methods \cite{LFCNN2015,LFCNN2017,LF-DCNN,LFNet,LFSSR,resLF,LF-InterNet} do not explicitly address the disparity issue. Instead, they use cascaded convolutions to achieve a large receptive field to cover the disparity range. As demonstrated in \cite{SOFVSR18,SOFVSR20}, it is difficult for SR networks to learn the non-linear mapping between LR and HR images under complex motion patterns. Consequently, the misalignment impedes the incorporation of angular information and leads to performance degradation. Therefore, specific mechanisms should be designed to handle the disparity problem in LF image SR.

  Inspired by the success of deformable convolution \cite{deformable,deformableV2} in video SR \cite{TDAN,EDVR,DNLN,ZoomSlowMo,D3Dnet}, in this paper, we propose a deformable convolution network (namely, LF-DFnet) to handle the disparity problem for LF image SR. Specifically, we design an angular deformable alignment module (ADAM) and a collect-and-distribute approach to achieve feature-level alignment and angular information incorporation. In ADAM, all side-view features are first aligned with the center-view feature to achieve feature collection. These collected features are then fused and distributed to their corresponding views by performing alignment with their original features. Through feature collection and distribution, angular information can be incorporated and encoded into each view. Consequently, the SR performance is evenly improved among different views. Moreover, we develop a novel LF dataset named NUDT to evaluate the performance of LF image SR methods under different disparity variations. All scenes in our NUDT dataset are rendered using the 3dsMax software\footnote{https://www.autodesk.eu/products/3ds-max/overview} and the baseline of virtual camera arrays is adjustable. In summary, the main contributions of this paper are as follows:
 \begin{itemize}
 \item We propose an LF-DFnet to achieve the state-of-the-art LF image SR performance by addressing the disparity problem.
 \item We propose an angular deformable alignment module and a collect-and-distribute approach to achieve high-quality reconstruction of each LF image. Compared to  \cite{resLF}, our approach avoids repetitive feature extraction and can exploit angular information from all SAIs.
 \item We develop a novel NUDT dataset by rendering synthetic scenes with adjustable camera baselines. Experiments on the NUDT dataset have demonstrated the robustness of our method with respect to disparity variations.
 \end{itemize}

The rest of this paper is organized as follows: In Secion~\ref{secRelated}, we briefly review the related work. In Section~\ref{secMethod}, we introduce the architecture of our LF-DFnet in details. In Section~\ref{secDataset}, we introduce our self-developed dataset. Experimental results are presented in Section~\ref{secExperiments}. Finally, we conclude this paper in Section~\ref{secConclusion}.

\section{Related Work}\label{secRelated}

 In this section, we briefly review the major works in single image SR (SISR), LF image SR, and deformable convolution.

\subsection{Single Image SR}

 The task of SISR is to generate a clear HR image from its blurry LR counterpart. Since an input LR image can be associated to multiple HR outputs, SISR is a highly ill-posed problem. Recently, several surveys \cite{anwar2020deep,yang2019deep,wang2020survey} have been published to comprehensively review SISR methods. Here, we only describe several mile-stone works in literature.

 Since Dong et al. \cite{SRCNN2014,SRCNN2015} proposed the seminal work of CNN-based SISR method SRCNN, deep learning-based methods have dominated this area due to their remarkable performance in terms of both accuracy and efficiency. By far, various networks have been proposed to continuously improve the SISR performance. Kim et al. \cite{VDSR} proposed a very deep SR network (i.e., VDSR) and achieved a significant performance improvement over SRCNN. Lim et al. \cite{EDSR} proposed an enhanced deep SR network (i.e., EDSR). With the combination of local and global residual connections, EDSR won the NTIRE 2017 SISR challenge \cite{NTIRE2017}. Zhang et al. \cite{RDN,zhang2020residual} proposed a residual dense network (i.e., RDN), which achieved a further improvement over the state-of-the-arts at that time. Subsequently, Zhang et al. \cite{RCAN} proposed a residual channel attention network (i.e., RCAN) by introducing a channel attention module and a residual in residual mechanism. More recently, Dai et al. \cite{SAN} proposed SAN by applying the second-order attention mechanism to SISR. Note that, RCAN \cite{RCAN} and SAN \cite{SAN} achieve the state-of-the-art SISR performance to date in terms of PSNR and SSIM.

 In summary, SISR networks are becoming increasingly deep and complicated, resulting in continuously improved capability in spatial information exploitation. Note that, performing SISR on LF images is a straightforward scheme to achieve LF image SR. However, the angular information is discarded in this scheme, resulting in limited performance.

\subsection{LF image SR}

  In the area of LF image SR, both traditional and deep learning-based methods are widely used. For traditional methods, various models have been developed for problem formulation. Wanner et al. \cite{wanner2013variational} proposed a variational method for LF image SR based on the estimated depth information. Mitra et al. \cite{mitra2012light} encoded LF structure via a Gaussian mixture model to achieve depth estimation, view synthesis, and LF image SR. Farrugia et al. \cite{farrugia2017super} decomposed HR-LR patches into subspaces and proposed a linear subspace projection method for LF image SR. Alain et al. proposed LFBM5D for LF image denoising \cite{alain2017light} and LF image SR \cite{LFBM5D} by extending BM3D filtering \cite{BM3D} to LFs. Rossi et al.  \cite{GB} developed a graph-based method to achieve LF image SR via graph optimization. Although the LF structure is well encoded by these models \cite{wanner2013variational,mitra2012light,farrugia2017super,LFBM5D,GB}, the spatial information cannot be fully exploited due to the poor representation capability of these handcrafted image priors.

  Recently, deep learning based SISR methods are demonstrated superior to traditional methods in spatial information exploitation. Inspired by these works, recent LF image SR methods adopted deep CNN to improve their performance. In the pioneering work LFCNN \cite{LFCNN2015,LFCNN2017}, SAIs were first separately super-resolved via SRCNN, and then fine-tuned in pairs to enhance both spatial and angular resolution. Subsequently, Yuan et al. \cite{LF-DCNN} proposed LF-DCNN to improve LFCNN by super-resolving each SAI via a more powerful SISR network EDSR and fine-tuning the initial results using a specially designed EPI-enhancement network. Apart from these two-stage SR methods, a number of one-stage network architectures have been designed for LF image SR. Wang et al. proposed a bidirectional recurrent network LFNet \cite{LFNet} by extending BRCN \cite{BRCN} to LFs. Zhang et al. \cite{resLF} proposed a multi-stream residual network  resLF by stacking SAIs along different angular directions as inputs to super-resolve the center-view SAI. Yeung et al. \cite{LFSSR} proposed LFSSR to alternately shuffle LF features between SAI pattern and macro-pixel image pattern for convolution. More recently, Jin et al. \cite{ATO} proposed an all-to-one LF image SR method (i.e., LF-ATO) and performed structural consistency regularization to preserve the parallax structure among reconstructed views. Wang et al. \cite{LF-InterNet} proposed an LF-InterNet to interact spatial and angular information for LF image SR. LF-ATO and LF-InterNet are state-of-the-art LF image SR methods to date and can achieve a high reconstruction accuracy.

  Although the performance is continuously improved by recent networks, the disparity problem has not been well addressed in literature. Several methods \cite{LFCNN2015,LFCNN2017,LF-DCNN,resLF} use stacked SAIs as their inputs, making pixels of same objects vary in spatial locations. In LFSSR \cite{LFSSR} and LF-InterNet \cite{LF-InterNet},  LF features are organized into a macro-pixel image pattern to incorporate angular information. However, pixels can fall into different macro-pixels due to the disparity problem. In summary, due to the lack of the disparity handling mechanism, the performance of these methods degrade when handling scenes with large disparities. Note that, LFNet \cite{LFNet} achieves LF image SR in a video SR framework and implicitly addresses the disparity issue via recurrent networks. Although all angular views can contribute to the final SR performance, the recurrent mechanism in LFNet \cite{LFNet} only takes SAIs from the same row or column as its inputs. Therefore, the angular information in LFs cannot be efficiently used.

\subsection{Deformable Convolution}

 The fixed kernel configuration in regular CNNs hinders the exploitation of long-range information. To address this problem, Dai et al. \cite{deformable} proposed deformable convolution by introducing additional offsets, which can be learned adaptively to make the convolution kernel process feature far away from its local neighborhood. Deformable convolutions have been applied to both high-level vision tasks \cite{bertasius2018object,deformable,zhao2018trajectory,sun2018integral}, and low-level vision tasks such as video SR \cite{TDAN,EDVR,DNLN,ZoomSlowMo}. Specifically, Tian et al. \cite{TDAN} proposed a temporal deformable alignment network (i.e., TDAN) by applying deformable convolution to align input video frames without explicit motion estimation. Wang et al. \cite{EDVR} proposed an enhanced deformable video restoration network (i.e., EDVR) by introducing a pyramid, cascading and deformable alignment module to handle large motions between frames. EDVR won the NTIRE19 video restoration and enhancement challenges \cite{NTIRE2019}. More recently, deformable convolution is integrated with non-local operation \cite{DNLN}, convolutional LSTM \cite{ZoomSlowMo} and 3D convolutions \cite{D3Dnet} to further enhance the video SR performance.

 In summary, existing deformable convolution-based video SR methods \cite{TDAN,EDVR,DNLN,ZoomSlowMo,D3Dnet} only perform unidirectional alignments to align neighborhood frames to the reference frame. However, in LF image SR, it is computational expensive to repetitively perform unidirectional alignments for each view to super-resolve all LF images. Consequently, we propose a collect-and-distribute approach to achieve bidirectional alignments using deformable convolutions. To the best of our knowledge, this is the first work to apply deformable convolutions to LF image SR.

\section{Network Architecture}\label{secMethod}

 In this section, we introduce our LF-DFnet in details. Following \cite{resLF, LFNet, LFSSR, LF-DCNN,LF-InterNet}, we convert input images from RGB channel space to YCbCr channel space and only super-resolve the Y channel images, leaving Cb and Cr channel images being bicubicly upscaled. Consequently, without considering the channel dimension, an LF can be formulated as a 4D tensor $\mathcal{L}\in\mathbb{R^{\mathrm{\mathit{U}\times\mathit{V}\times\mathit{H}\times\mathit{W}}}}$, where $U$ and $V$ represent angular dimensions $H$ and $W$ represent spatial dimensions. Specifically, a 4D LF can be considered as a  $U\times V$ array of SAIs, and the resolution of each SAI is $H\times W$. Following  \cite{resLF, LFNet, LFSSR, LF-DCNN,LF-InterNet}, we achieve LF image SR using SAIs distributed in a square array, i.e.,  $\mathit{U}=\mathit{V}=\mathit{A}$.

\begin{figure*}[t]
\centering
\includegraphics[width=18cm]{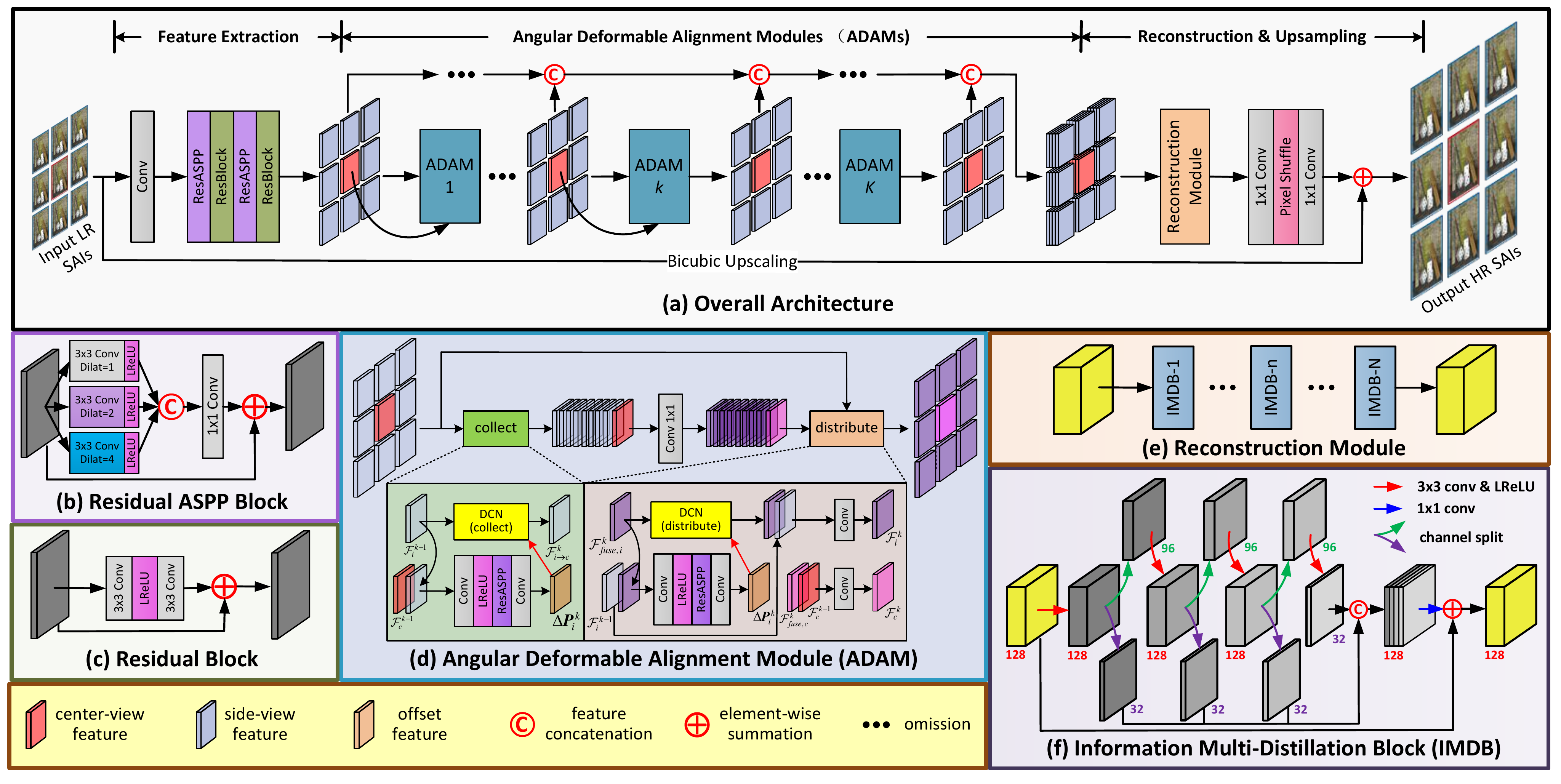}
\caption{{An overview of our LF-DFnet.}
\label{LF-DFnet}}
\end{figure*}

 As illustrated in Fig.~\ref{LF-DFnet}(a), our LF-DFnet takes LR SAIs as its inputs and sequentially performs feature extraction (Section \ref{FEmodule}), angular deformable alignment (Section \ref{DFmodule}), reconstruction and upsampling (Section \ref{RCUSmodule}).

\subsection{Feature Extraction Module}\label{FEmodule}

 Discriminative feature representation with rich spatial context information is beneficial to the subsequent feature alignment and SR reconstruction steps. Therefore, a large receptive field with a dense pixel sampling rate is required to extract hierarchical features. To this end, we follow \cite{PASSRnet} and use residual atrous spatial pyramid pooling (ASPP) module as the feature extraction module in our LF-DFnet.

 As shown in Fig.~\ref{LF-DFnet}(a), input SAIs are first processed by a $1\times1$ convolution to generate initial features, and then fed to residual ASPP modules (Fig.~\ref{LF-DFnet}(b)) and residual blocks (Fig.~\ref{LF-DFnet}(c)) for deep feature extraction. Note that, each view is processed separately and the weights in our feature extraction module are shared among these views. In each residual ASPP block, three $3\times3$ dilated convolutions (with dilation rates of 1, 2, 4, respectively) are combined in parallel to extract hierarchical features with dense sampling rates. After activation with a Leaky ReLU layer (with a leaky factor of 0.1), features of these three branches are concatenated and fused by a $1\times1$ convolution. Finally, both the center-view feature $\mathcal{F}_{c}\in\mathbb{R^{\mathrm{\mathit{H}\times\mathit{W}\times\mathit{C}}}}$ and side-view features $\mathcal{F}_{i}\in\mathbb{R^{\mathrm{\mathit{H}\times\mathit{W}\times\mathit{C}}}}\,(i=1, 2,\cdots, A^{2}-1)$ are generated by our feature extraction module.
 Following \cite{resLF}, we set the feature depth to 32 (i.e., $C=32$). The effectiveness of residual ASPP module is demonstrated in Section~\ref{secAblation}.

\subsection{Angular Deformable Alignment Module (ADAM)}\label{DFmodule}

 Given features generated by the feature extraction module, the main objective of ADAM is to perform alignment between the center-view feature and each side-view feature. Here, we propose a bidirectional alignment approach (i.e., collect-and-distribute) to incorporate angular information. Specifically, side-view features are first aligned with the center-view feature to perform feature collection. Then, these aligned features are fused by a $1\times1$ convolution to incorporate angular information. Afterwards, the fused features are further aligned with their original features to achieve feature distribution. In this way, angular information can be jointly incorporated into each angular view, and the SR performance of all perspectives can be evenly improved (see Section~\ref{secAblation}). In this paper, we cascade $K$ ADAMs to perform feature collection and feature distribution. Without loss of generality, we take the $k^{\mathrm{th}}\, (k=1,2,\cdots,K)$ ADAM as an example to introduce its mechanism, as shown in Fig.~\ref{LF-DFnet}(d).

 The core component of ADAM is deformable convolution, which is used to align features according to their corresponding offsets. In our implementation, we use a deformable convolution for feature collection and another deformable convolution with shared weights for feature distribution. The first deformable convolution, which is used for feature collection, takes the $(k-1)^{\mathrm{th}}$ side-view feature $\mathcal{F}_{i}^{k-1}$ and learnable offsets $\Delta \bm{P}_{i\rightarrow c}^{k}$ as its input to generate $\mathcal{F}_{i\rightarrow c}^{k}$ (which is aligned to the center view). That is,
 \begin{equation}
\mathcal{F}_{i\rightarrow c}^{k} = H_{dcn}^{k}\left(\mathcal{F}_{i}^{k-1}, \Delta \bm{P}_{i\rightarrow c}^{k}\right),
\end{equation}
 where $H_{dcn}^{k}$ represents the deformable convolution in the $k^{\mathrm{th}}$ deformable block, $\Delta\bm{P}_{i\rightarrow c}^{k}=\{\Delta \bm{p}_n\} \in\mathbb{R}^{\mathit{H}\times\mathit{W}\times\mathit{C'}}$ is the offset of $\mathcal{F}_{i}^{k-1}$ with respect to $\mathcal{F}_{c}$. More specifically, for each position $\bm{p}_0=(x_0, y_0)$ on $\mathcal{F}_{i\rightarrow c}^{k}$, we have
\begin{equation}
\mathcal{F}_{i\rightarrow c}^{k}(\bm{p}_0) = \sum \limits_{\bm{p}_{n}\in \mathbf{R}} w(\bm{p}_n)\cdot\mathcal{F}_{i}^{k-1} \left(\bm{p}_0 + \bm{p}_n + \Delta \bm{p}_n \right),
\end{equation}
 where $\bm{R}=\{(-1,-1),(-1,0),\cdots,(0,1),(1,1)\}$ represents a $3\times3$ neighborhood region centered at $\bm{p}_0$. $\bm{p}_n\in\bm{R}$ is the predefined integral offset. $\Delta \bm{p}_n$ is an additional learnable offset, which is added to the predefined offset $\bm{p}_n$ to make the positions of deformable kernels spatially-variant. Thus, information far away from $\bm{p}_0$ can be adaptively processed by deformable convolution. Since $\Delta \bm{p}_n$ can be fractional, we follow \cite{deformable} to perform bilinear interpolation to generate exact offset values in our implementation.

 Since an accurate offset is beneficial to deformable alignment, we design an offset generation branch to learn offset $\Delta \bm{P}_{i\rightarrow c}^{k}$ in Eq.~(1). As illustrated in Fig.~\ref{LF-DFnet}(d), the side-view feature $\mathcal{F}_{i}^{k-1}$ is first concatenated with the center-view feature $\mathcal{F}_{c}$, and then fed to a $1\times1$ convolution for feature depth reduction. To handle the complicated and large motions between $\mathcal{F}_{i}^{k-1}$ and $\mathcal{F}_{c}$, a residual ASPP module (which is identical to that in the feature extraction module but with different weights) is applied to enlarge the receptive field while maintaining a dense sampling rate. The residual ASPP module enhances the exploitation of angular dependencies between the center view and side views, resulting in improved SR performance (see Section~\ref{secAblation}). Finally, another $1\times1$ convolution with $C'=18$ output channels is used to generate the offset feature.

 Once all side-view features are aligned to the center view, a $1\times1$ convolution is performed to the collected features to fuse the complementary angular information. That is,
\begin{equation}
\mathcal{F}_{fuse}^{k} = H_{fuse}^{k}\left(\mathcal{F}_{collect}^{k}\right),
\end{equation}
 where $\mathcal{F}_{collect}^{k} =\left[\mathcal{F}_{1\rightarrow c}^{k}, \mathcal{F}_{2\rightarrow c}^{k}, \cdots,\mathcal{F}_{(A^{2}-1)\rightarrow c}^{k},\mathcal{F}_{c}^{k-1}\right]$ denotes the concatenation of the center-view feature and all the aligned side-view features, $H_{fuse}^{k}$ denotes a $1\times1$ convolution for angular information incorporation. Finally, $\mathcal{F}_{fuse}^{k}\in\mathbb{R^{\mathrm{\mathit{H}\times\mathit{W}\times\mathit{A^{2}C}}}}$ is obtained by performing the above fusion operation.

 To super-resolve all LF images, the incorporated angular information in $\mathcal{F}_{fuse}^{k}$ need to be propagated into each side view. Consequently, we apply deformable convolution for the second time to perform feature distribution. Specifically, the fused feature $\mathcal{F}_{fuse}^{k}$ is first split along the channel dimension to generate $A^2$ sub-features. Then, for a specific side view $i$, a sub-feature $\mathcal{F}_{fuse,i}^{k}\in\mathbb{R^{\mathrm{\mathit{H}\times\mathit{W}\times\mathit{C}}}}$ is concatenated with the corresponding original side-view feature $\mathcal{F}_{i}^{k-1}$ and fed to the offset generation branch to produce offset $\Delta \bar{\bm{P}}_{i}^{k}$ ($\mathcal{F}_{fuse,i}^{k}$ with respect to $\mathcal{F}_{i}^{k-1}$). The final distributed feature of side view $i$ can be obtained according to
  \begin{equation}
 \mathcal{F}_{i}^{k} = H_{squeeze}^{k} \left( \left[ H_{dcn}^{k} \left( \mathcal{F}_{fuse,i}^{k}, \Delta \bar{\bm{P}}_{i}^{k}\right), \mathcal{F}_{i}^{k-1} \right] \right),
 \end{equation}
where $H_{squeeze}^{k}$ denotes a $1\times1$ convolution to reduce the channel numbers from 2$C$ to $C$. Note that, the weights of both the offset generation branch and deformable convolution are shared between feature collection and feature distribution. The center-view feature is updated according to
  \begin{equation}
 \mathcal{F}_{c}^{k} = H_{squeeze}^{k} \left( \left[\mathcal{F}_{fuse,c}^{k}, \mathcal{F}_{c}^{k-1} \right] \right).
 \end{equation}

 After feature distribution, both the center-view feature $\mathcal{F}_{c}^{k}$ and side-view features $\mathcal{F}_{i}^{k},(i=1, 2,\cdots, A^{2}-1)$ are produced by the $k^{\mathrm{th}}$ ADAM. In this paper, we cascade three ADAMs (i.e., $K=3$) to achieve repetitive feature collection and distribution. Consequently, angular information can be repetitively incorporated into the center view and then propagated to all side views, resulting in notable performance improvements (see Section~\ref{secAblation}).

 \begin{figure}
 \centering
 \includegraphics[width=8.8cm]{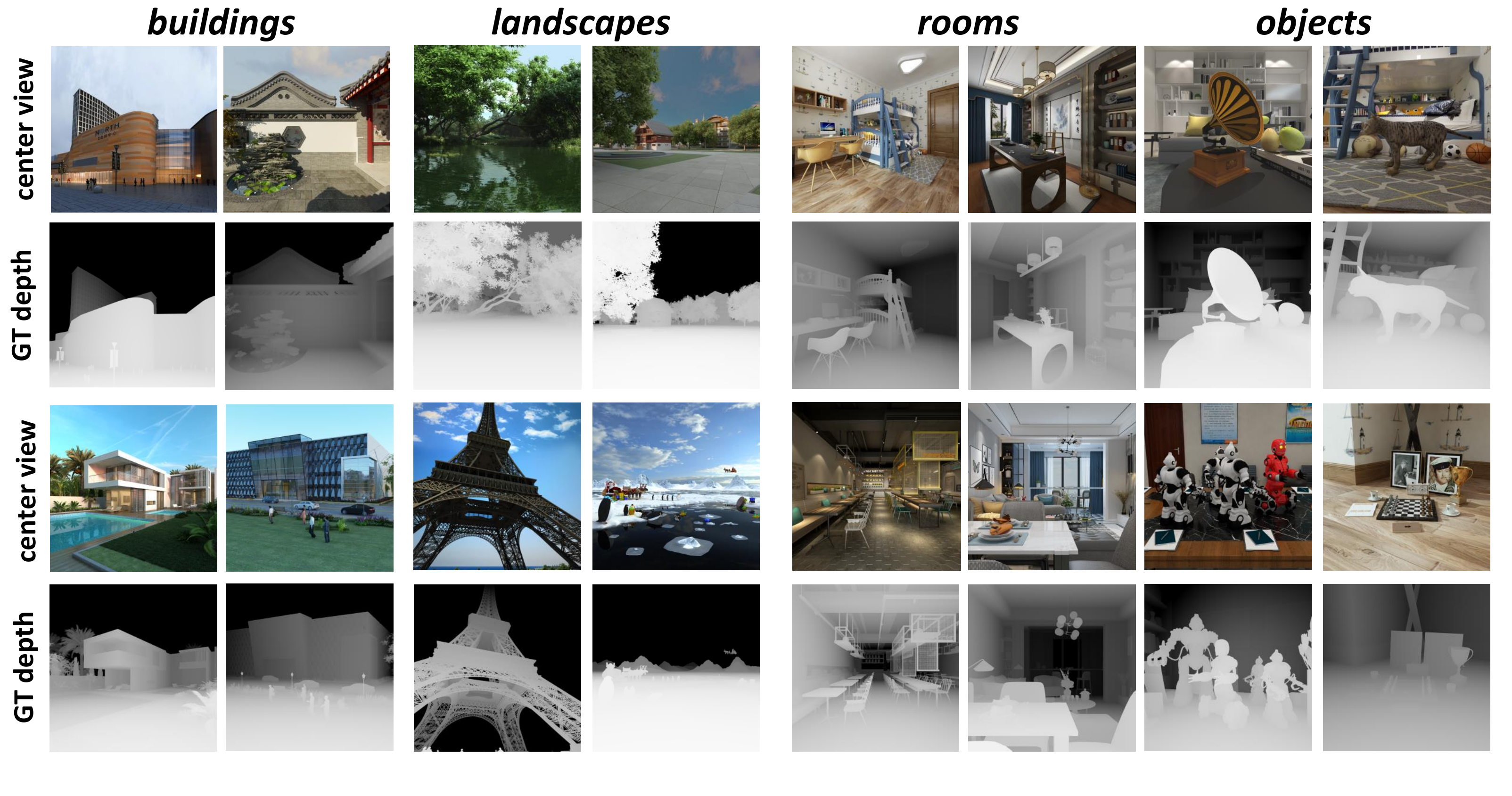}
 \caption{{Example images and their groundtruth depth maps in our NUDT dataset.}
 \label{NUDT}}
 \end{figure}
\subsection{Reconstruction \& Upsampling Module}\label{RCUSmodule}

 To achieve high reconstruction accuracy, the spatial and angular information has to be incorporated. Since preceding modules in our LF-DFnet have produced angular-aligned hierarchical features, a reconstruction module is needed to fuse these features for LF image SR. Following \cite{IMDN}, we propose a reconstruction module with information multi-distillation blocks (IMDB). By adopting distillation mechanism to gradually extract and process hierarchical features, superior SR performance can be achieved with a small number of parameters and a low computational cost \cite{AIM2019}.

 The overall architecture of our reconstruction module is illustrated in Fig.~\ref{LF-DFnet}(e). For each view, the outputs of the feature extraction module and each ADAM are concatenated and fed to several stacked IMDBs. The structure of IMDB is illustrated in Fig.~\ref{LF-DFnet}(f). Specifically, in each IMDB, the input feature is first processed by a $3\times3$ convolution and a Leaky ReLU layer. The processed feature is then split into two parts along the channel dimension, resulting in a narrow feature (with 32 channels) and a wide feature (with 96 channels). The narrow feature is preserved and directly fed to the final bottleneck of this IMDB, while the wide feature is fed to a $3\times3$ convolution to enlarge its channels to 128 for further refinement. In this way, useful information can be gradually distilled, and the SR performance is improved in an efficient manner. Finally, features of different stages in the IMDB are concatenated and processed by a $1\times1$ convolution for local residual learning.

 Features obtained from the reconstruction module are finally fed to an upsampling module. Specifically, a $1\times1$ convolution is first applied to the reconstructed features to extend their depth to $\alpha^{2}C$, where $\alpha$ is the upsampling factor. Then, pixel shuffle is performed to upscale the reconstructed feature to the target resolution $\alpha H \times \alpha W$. Finally, a $1\times1$ convolution is applied to squeeze the number of feature channels to 1 to generate super-resolved SAIs.

 \begin{figure}
 \centering
 \includegraphics[width=8.8cm]{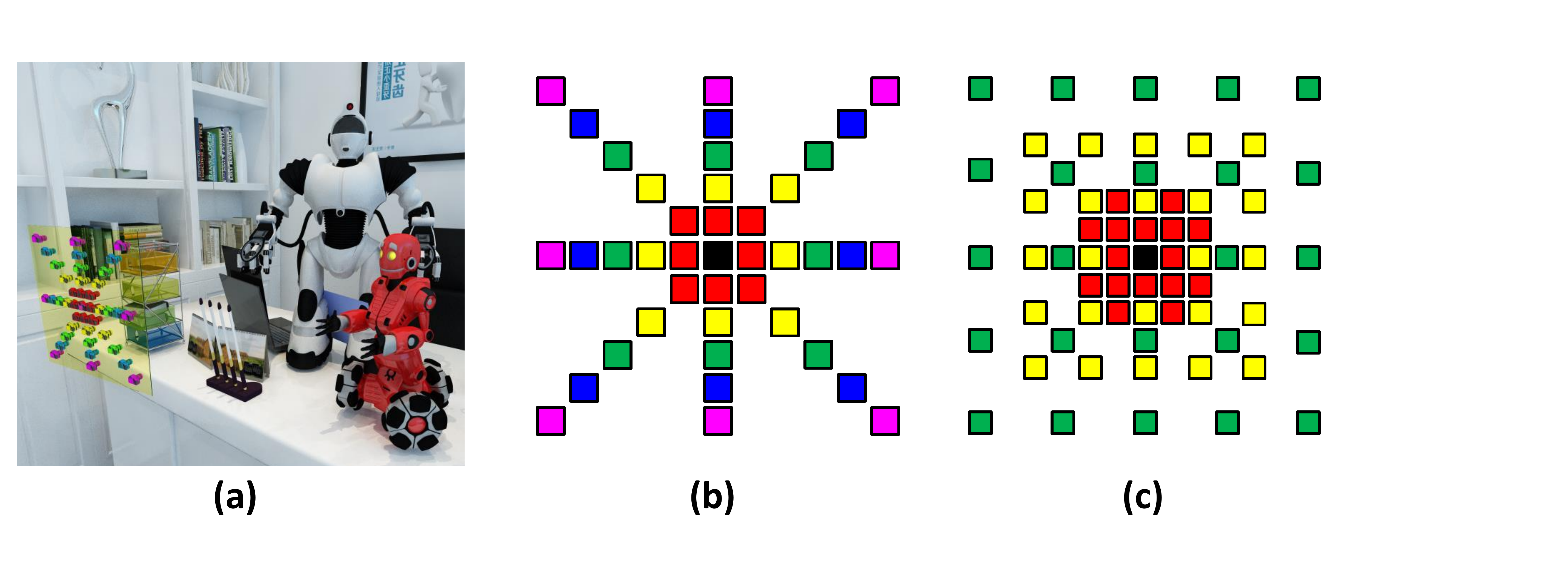}
 \caption{{An illustration of the concentric configuration. (a) Configuration of scene \textit{Robots}. Here, $3\times3$ camera arrays of 5 different settings of baselines are used as examples. Blocks on the translucent yellow plane denote virtual cameras, where camera arrays of different baselines are drawn in different colors. (b) $3\times3$ concentric configuration with 5 different settings of baselines. (c) $5\times5$ concentric configuration with 3 different settings of baselines.}
 \label{figBaseline}}
 \end{figure}

\begin{table*}
 \centering
 \scriptsize
 \renewcommand\arraystretch{1.2}
 \caption{Main characteristics of several popular LF datasets. Note that, average scores are reported for spatial resolution (SpaRes), single-image perceptual quality metrics (i.e., BRISQUE \cite{BRISQUE}, NIQE \cite{NIQE}, CEIQ \cite{CEIQ}, ENIQA \cite{ENIQA})  and LF quality assessment metrics (i.e., NRLFQA \cite{NRLFQA}).}\label{DatasetCompare}
 \begin{tabular}{|l|c|c|c|c|c|c|c|c|c|c|}
 \hline
 Datasets & Type &  \#Scenes & AngRes & SpaRes ($\uparrow$) & GT Depth & BRISQUE ($\downarrow$) & NIQE  ($\downarrow$)  & CEIQ ($\uparrow$) & ENIQA  ($\downarrow$) & NRLFQA ($\uparrow$)\\[0.5pt]
 \hline
 EPFL \cite{EPFL}              &  real (lytro)   & 119 &14$\times$14&  0.034 Mpx  & \ding{55} & 47.19 & 5.820 & 3.286 & 0.212 & 2.970\\
 HCInew \cite{HCInew}       & synthetic      & 24   &9$\times$9& 0.026 Mpx  & \ding{51} & \underline{14.80} & 3.833 & 3.153 & \underline{0.087} & 2.915\\
 HCIold \cite{HCIold}         & synthetic      & 12   &9$\times$9&  0.070 Mpx  & \ding{51} & 24.17 & \textbf{2.985} & \underline{3.369} & 0.117 & 2.720\\
 INRIA \cite{INRIA}           & real (lytro)    & 57   &14$\times$14&  0.027 Mpx  & \ding{55} & 23.56 & 5.338 & 3.184 &  0.160 & \underline{2.983}\\
 STFgantry \cite{STFgantry} & real (gantry) & 12   &17$\times$17&  \textbf{0.118} Mpx & \ding{55}  & 25.28 & 4.246 & 2.781 &  0.232 & \textbf{2.993}\\
 NUDT (Ours) & synthetic & 32 &9$\times$9& \underline{0.105} Mpx & \ding{51} & \textbf{8.901} & \underline{3.593} & \textbf{3.375} &  \textbf{0.041} & \underline{2.983}\\
 \hline
 \end{tabular}
 \leftline{
 \begin{tabular}{l}
 ~~Note: 1) Mpx denotes mega-pixels per image. ~2) The best results are in \textbf{bold} faces and the second best results are \underline{underlined}. ~3) Lower scores of BRISQUE, NIQE,\\ ~~~~~~~~~~ENIQA and higher scores of CEIQ, NRLFQA indicate better quality.
 \end{tabular}}
 \end{table*}

\section{The \textit{NUDT} Dataset}\label{secDataset}
 LF images captured by different devices (especially camera arrays) usually have significantly different baseline lengths. It is therefore, necessary to know how existing LF algorithms work under baseline variations, including those developed for depth estimation \cite{EPINET,park2017robust,lee2019complex,jeon2018depth,sheng2018occlusion,SPO}, view synthesis \cite{wu2017light,wu2018light,wu2019learning,vagharshakyan2017light,wing2018fast,wang2018end,zhang2017micro,meng2019high,jin2020learning}, and image SR \cite{meng2020high,ATO,farrugia2019light,farrugia2020simple,gul2018spatial}. However, all existing LF datasets  \cite{EPFL,HCInew,HCIold,INRIA,STFgantry} only include images with fixed baselines. To this end, we introduce a novel LF dataset (namely, the NUDT dataset) with adjustable baselines to facilitate the study of LF algorithms under baseline variations.

 \subsection{Technical Details}
 Our NUDT dataset has 32 synthetic scenes and covers diverse scenarios (see Fig.~\ref{NUDT}).  All scenes in our dataset are rendered using the 3dsMax software\footnote{https://www.autodesk.eu/products/3ds-max/overview}, and have an angular resolution of $9\times9$ and a spatial resolution of $1024\times1024$. Groundtruth depth maps are available for LF depth/disparity estimation methods. During the image rendering process,  all virtual cameras in the array have identical internal parameters and are coplanar with the parallel optical axes. To capture LF images with different baselines, we used a concentric configuration to align camera arrays at the center views. In this way, LF images of different baselines share the same center-view SAI and groundtruth depth map. An illustration of our concentric configuration is shown in Fig.~\ref{figBaseline}. For each scene, we rendered LF images with 5 different baselines. Note that, we tuned the parameters (e.g., lighting, and depth range) to better reflect real scenes. Consequently, our dataset has a high perceptual quality, which will be introduced in the next subsection.

 \subsection{Comparison to Existing Datasets}
 In this section, we compare our NUDT dataset to several popular LF datasets \cite{EPFL,HCInew,HCIold,INRIA,STFgantry}. Following \cite{Flickr1024}, we use four no-reference image quality assessment (NRIQA) metrics (i.e, BRISQUE \cite{BRISQUE}, NIQE \cite{NIQE}, CEIQ \cite{CEIQ}, ENIQA \cite{ENIQA}) to evaluate the perceptual quality of the center-view images of these datasets. Besides, we also use a no-reference LF quality assessment metric (i.e., NRLFQA \cite{NRLFQA}) to evaluate the spatial quality and angular consistency of LFs. As shown in Table~\ref{DatasetCompare}, our NUDT dataset achieves the best scores in BRISQUE, CEIQ, and ENIQA, and achieves the second best scores in NIQE and NRLFQA. That is, LF images in our NUDT dataset are angular consistent and have high perceptual quality. Meanwhile, our dataset has more scenes (see \#Scenes)  and higher image resolution (see SpaRes) than the synthetic HCInew \cite{HCInew} and HCIold \cite{HCIold} datasets.

\begin{table}
\centering
\renewcommand\arraystretch{1.1}
\caption{Public datasets used in our experiments.}\label{datasets}
\begin{tabular}{|l|c|c|}
\hline
Datasets~~~~~~~~~~~~&~~~~~\#Training~~~~~&~~~~~\#Test~~~~~\\[0.5pt]
\hline
EPFL \cite{EPFL}       &  70  & 10 \\
HCInew \cite{HCInew}       &  20  & 4  \\
HCIold \cite{HCIold}       &  10  & 2  \\
INRIA \cite{INRIA}         &  35  & 5  \\
STFgantry \cite{STFgantry} &  9   & 2  \\
\hline
Total                      &  \textbf{144} & \textbf{23} \\
\hline
\end{tabular}
\vspace{0cm}
\end{table}

\section{Experiments}\label{secExperiments}
In this section, we first introduce our implementation details. Then, we compare our LF-DFnet to several state-of-the-art SISR and LF image SR methods from different perspectives. Finally, we present ablation studies to investigate our network.

\begin{table*}
\caption{
PSNR$/$SSIM$/$PI values achieved by different methods for $2\times$ and $4\times$SR. For PSNR and SSIM, larger values indicate higher reconstruction quality. For PI \cite{Pirm2018}, smaller values indicate higher perceptual quality. The best results are in \textcolor{red}{red} and the second best results are in \textcolor{blue}{blue}.}\label{tabComparison}
\centering
\renewcommand\arraystretch{1.1}
\begin{tabular}{|l|c|lllll|}
\hline
\multirow{2}*{Method}&\multirow{2}*{Scale}&  \multicolumn{5}{c|}{Dataset} \\
\cline{3-7}
  &  & EPFL \cite{EPFL} &HCInew \cite{HCInew} & HCIold \cite{HCIold} & INRIA \cite{INRIA} & STFgantry \cite{STFgantry} \\
\hline
Bicubic                                     & $2\times$  & 29.50$/$0.9350$/$5.633~~~& 31.69$/$0.9335$/$5.141~~~& 37.46$/$0.9776$/$5.715~~~& 31.10$/$0.9563$/$5.592~~~& 30.82$/$0.9473$/$6.058\\
VDSR \cite{VDSR}                  & $2\times$ & 32.50$/$0.9599$/$4.874 & 34.37$/$0.9563$/$4.080 & 40.61$/$0.9867$/$4.211 & 34.43$/$0.9742$/$4.636 & 35.54$/$0.9790$/$4.791\\
EDSR \cite{EDSR}                   & $2\times$ & 33.09$/$0.9631$/$4.749 & 34.83$/$0.9594$/$3.914 & 41.01$/$0.9875$/$4.051 & 34.97$/$0.9765$/$4.568 & 36.29$/$0.9819$/$4.642\\
RCAN \cite{RCAN}                 & $2\times$ & 33.16$/$0.9635$/$4.780 & 34.98$/$0.9602$/$3.940 & 41.05$/$0.9875$/$4.063 & 35.01$/$0.9769$/$4.591 & 36.33$/$0.9825$/$4.652\\
LFBM5D \cite{LFBM5D}         & $2\times$ & 31.15$/$0.9545$/$4.965 & 33.72$/$0.9548$/$4.525 & 39.62$/$0.9854$/$4.755 & 32.85$/$0.9695$/$4.998 & 33.55$/$0.9718$/$5.159\\
GB \cite{GB}                           & $2\times$ & 31.22$/$0.9591$/$4.644 & 35.25$/$0.9692$/$3.741 & 40.21$/$0.9879$/$4.105 & 32.76$/$0.9724$/$4.461 & 35.44$/$0.9835$/$\textcolor{red}{4.469}\\
resLF \cite{resLF}                   & $2\times$ & 32.75$/$0.9672$/$\textcolor{red}{4.480} & 36.07$/$0.9715$/$3.678 & 42.61$/$0.9922$/$3.818 & 34.57$/$0.9784$/$\textcolor{blue}{4.365} & 36.89$/$0.9873$/$4.580\\
LFSSR \cite{LFSSR}              & $2\times$ & 33.69$/$0.9748$/$4.623 & 36.86$/$0.9753$/$3.702 & 43.75$/$0.9939$/$3.755 & 35.27$/$0.9834$/$4.504 & 38.07$/$0.9902$/$4.631\\
LF-InterNet \cite{LF-InterNet}  & $2\times$ & \textcolor{blue}{34.14}$/$\textcolor{blue}{0.9761}$/$4.580 & \textcolor{blue}{37.28}$/$\textcolor{blue}{0.9769}$/$\textcolor{blue}{3.658} & \textcolor{red}{44.45}$/$\textcolor{red}{0.9945}$/$\textcolor{blue}{3.710} & \textcolor{blue}{35.80}$/$\textcolor{red}{0.9846}$/$4.484 & \textcolor{blue}{38.72}$/$\textcolor{blue}{0.9916}$/$4.602\\
LF-DFnet (Ours)                      & $2\times$ & \textcolor{red}{34.44}$/$\textcolor{red}{0.9766}$/$\textcolor{blue}{4.512} & \textcolor{red}{37.44}$/$\textcolor{red}{0.9786}$/$\textcolor{red}{3.623} & \textcolor{blue}{44.23}$/$\textcolor{blue}{0.9943}$/$\textcolor{red}{3.680} & \textcolor{red}{36.36}$/$\textcolor{blue}{0.9841}$/$\textcolor{red}{4.316} & \textcolor{red}{39.61}$/$\textcolor{red}{0.9935}$/$\textcolor{blue}{4.549}\\
\hline
Bicubic                                       & $4\times$  & 25.14$/$0.8311$/$7.802 & 27.61$/$0.8507$/$7.651 & 32.42$/$0.9335$/$7.644 & 26.82$/$0.8860$/$7.574 & 25.93$/$0.8431$/$7.531\\
VDSR \cite{VDSR}                   & $4\times$ & 27.25$/$0.8782$/$6.700 & 29.31$/$0.8828$/$6.417 & 34.81$/$0.9518$/$6.416 & 29.19$/$0.9208$/$6.679 & 28.51$/$0.9012$/$6.503\\
EDSR \cite{EDSR}                    & $4\times$ & 27.84$/$0.8858$/$6.293 & 29.60$/$0.8874$/$6.095 & 35.18$/$0.9538$/$6.311 & 29.66$/$0.9259$/$6.248 & 28.70$/$0.9075$/$5.923\\
RCAN \cite{RCAN}                  & $4\times$ & 27.88$/$0.8863$/$6.231 & 29.63$/$0.8880$/$5.991 & 35.20$/$0.9540$/$6.233 & 29.76$/$0.9273$/$6.196 & 28.90$/$0.9110$/$5.917\\
ESRGAN \cite{ESRGAN}        & $4\times$ & 24.35$/$0.7968$/$\textcolor{red}{3.852} & 26.20$/$0.8003$/$\textcolor{red}{3.309} & 30.69$/$0.9099$/$\textcolor{red}{3.317} & 26.49$/$0.8652$/$\textcolor{red}{3.794} & 25.46$/$0.8502$/$\textcolor{red}{4.342}\\
LFBM5D \cite{LFBM5D}        & $4\times$ & 26.61$/$0.8689$/$6.901 & 29.13$/$0.8823$/$6.534 & 34.23$/$0.9510$/$6.579 & 28.49$/$0.9137$/$6.888 & 28.30$/$0.9002$/$6.741\\
GB \cite{GB}                          & $4\times$ & 26.02$/$0.8628$/$7.217 & 28.92$/$0.8842$/$6.470 & 33.74$/$0.9497$/$6.641 & 27.73$/$0.9085$/$7.220 & 28.11$/$0.9014$/$6.648\\
resLF \cite{resLF}                  & $4\times$ & 27.46$/$0.8899$/$\textcolor{blue}{5.509} & 29.92$/$0.9011$/$\textcolor{blue}{5.109} & 36.12$/$0.9651$/$5.851 & 29.64$/$0.9339$/$\textcolor{blue}{5.615} & 28.99$/$0.9214$/$\textcolor{blue}{5.281}\\
LFSSR \cite{LFSSR}             & $4\times$ & 28.27$/$0.9080$/$5.899 & 30.72$/$0.9124$/$5.504 & 36.70$/$0.9690$/$5.875 & 30.31$/$0.9446$/$5.818 & 30.15$/$0.9385$/$5.815\\
LF-InterNet \cite{LF-InterNet}& $4\times$ & \textcolor{blue}{28.67}$/$\textcolor{blue}{0.9143}$/$6.061 & \textcolor{blue}{30.98}$/$\textcolor{blue}{0.9165}$/$5.594 & \textcolor{blue}{37.11}$/$\textcolor{blue}{0.9715}$/$5.844 & \textcolor{blue}{30.64}$/$\textcolor{blue}{0.9486}$/$5.919 & \textcolor{blue}{30.53}$/$\textcolor{blue}{0.9426}$/$5.867\\
LF-DFnet (Ours)                     & $4\times$ &\textcolor{red}{28.77}$/$\textcolor{red}{0.9165}$/$5.836 & \textcolor{red}{31.23}$/$\textcolor{red}{0.9196}$/$5.290 & \textcolor{red}{37.32}$/$\textcolor{red}{0.9718}$/$\textcolor{blue}{5.677} & \textcolor{red}{30.83}$/$\textcolor{red}{0.9503}$/$5.649 & \textcolor{red}{31.15}$/$\textcolor{red}{0.9494}$/$5.670\\
\hline
\end{tabular}
\end{table*}

\subsection{Implementation Details}\label{Implementation}

  As listed in Table~\ref{datasets}, we used 5 public LF datasets in our experiments for both training and test. For the LF datasets (i.e., EPFL \cite{EPFL} and INRIA \cite{INRIA}) recorded by Lytro cameras, we follow \cite{farrugia2017super,LFNet} to use the \textit{Light Field Toolbox v0.4} \cite{dansereau2013decoding} to decode raw LF images and extract 4D LF data.  All LFs in these datasets have an angular resolution of $9\times9$. In the training stage, we cropped each SAI into patches with a stride of 32, and used the bicubic downsampling approach to generate LR patches of size $32\times32$. We performed random horizontal flipping, vertical flipping, and 90-degree rotation to augment the training data by 8 times. Note that, both spatial and angular dimensions need to be flipped or rotated during data augmentation to maintain LF structures.

 By default, we used the model with $K=3$, $N=4$, $C=32$, and an angular resolution of $5\times5$ for both $2\times$ and $4\times$ SR. Our network was trained using the $L_1$ loss function and optimized using the Adam method \cite{Adam}. We initialized the weights and bias of the last convolution layer in the offset generation branch with zero values, and used the Kaiming method \cite{KaimingInit} to initialize other parts of the network. Our LF-DFnet was implemented in PyTorch on a PC with two NVidia RTX 2080Ti GPUs. The batch size was set to 8 and the learning rate was initially set to $2\times10^{-4}$ and decreased by a factor of 0.5 for every 15 epochs. The training was stopped after 50 epochs.

 We used PSNR, SSIM, and Perception Index (PI) \cite{Pirm2018} as quantitative metrics for performance evaluation. Note that, PSNR and SSIM were calculated on the Y channel images and PI was calculated on the RGB channel images. To obtain the metric score (e.g., PSNR) for a dataset with $M$ test scenes (each scene with an angular resolution of $A\times A$), we first calculated this metric on $A\times A$ SAIs on each scene separately, then obtained the score for each scene by averaging its $A^2$ scores, and finally obtained the score for this dataset by averaging the scores of all $M$ scenes.

\begin{figure*}
\centering
\includegraphics[width=18cm]{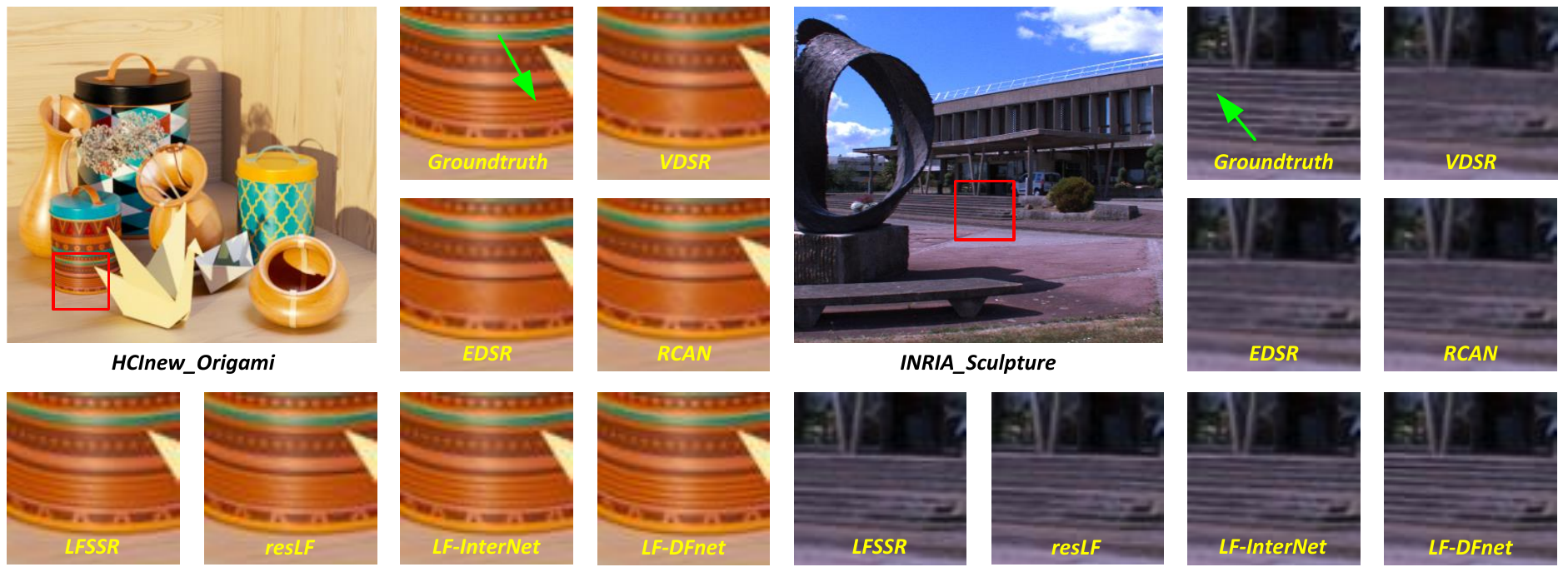}
\caption{Visual results of 2$\times$SR.} \label{2xSR}
\end{figure*}

\begin{figure*}[t]
\centering
\includegraphics[width=18cm]{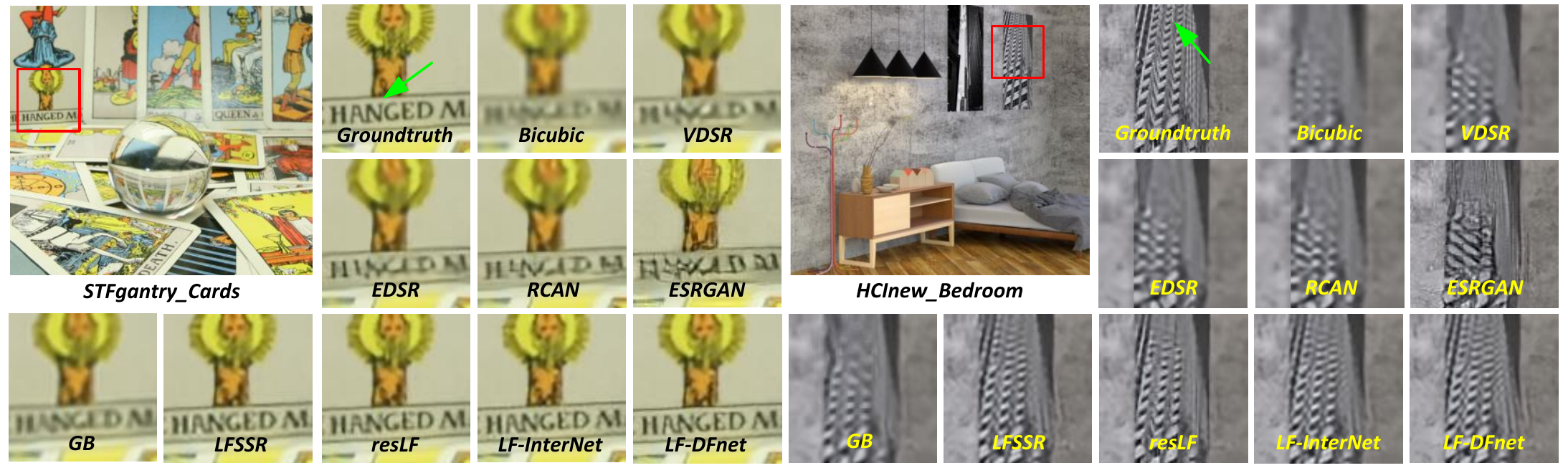}
\caption{Visual results of 4$\times$SR.} \label{4xSR}
\end{figure*}

\subsection{Comparison to the State-of-the-arts}\label{secComparison}

We compare our method to several state-of-the-art methods, including 4 single image SR methods (i.e., VDSR \cite{VDSR}, EDSR \cite{EDSR}, RCAN \cite{RCAN}, and ESRGAN \cite{ESRGAN}) and 5 LF image SR methods (i.e., LFBM5D \cite{LFBM5D}, GB \cite{GB}, LFSSR \cite{LFSSR}, resLF \cite{resLF}, and LF-InterNet \cite{LF-InterNet}). For fair comparison, we have retrained all deep-learning methods \cite{VDSR,EDSR,RCAN,ESRGAN,LFSSR,resLF,LF-InterNet} on the same training datasets as our LF-DFnet. We also include bicubic interpolation as a baseline method. For simplicity, we only present the results on $5\times5$ LFs for $2\times$ and $4\times$ SR.

\subsubsection{Quantitative Results}
Quantitative results are presented in Table \ref{tabComparison}. Our LF-DFnet achieves the highest PSNR and SSIM scores on all the 5 datasets for $4\times$SR and on 4 of 5 datasets (i.e., EPFL \cite{EPFL},  HCInew \cite{HCInew}, INRIA \cite{INRIA} and STFgantry \cite{STFgantry}) for $2\times$SR. In terms of the PI metric \cite{Pirm2018}, our method achieves the state-of-the-art performance for $2\times$SR, but is slightly inferior to ESRGAN \cite{ESRGAN} and resLF \cite{resLF} for $4\times$SR. Note that, PI is a no-reference metric for perceptual quality evaluation and cannot measure the faithfulness of resultant images. ESRGAN achieves the highest PI scores by generating clear but unfaithful textures (see Fig.~\ref{4xSR}). It is also worth noting that, the PSNR and SSIM improvements of our LF-DFnet are very significant on the STFgantry dataset for $2\times$SR. That is because, scenes in the STFgantry dataset are captured by a moving camera mounted on a gantry, and thus have relatively large baselines and significant disparity variations. Our LF-DFnet can handle this disparity problem by using deformable convolution for angular alignment, while maintaining promising performance for LFs with small baselines (e.g., LFs on the EPFL \cite{EPFL} and INRIA \cite{INRIA} datasets). More analyses with respect to different baseline lengths are presented in Section~\ref{secBaseline}.

\subsubsection{Qualitative Results}
 Qualitative results for $2\times$ and $4\times$ SR are shown in Figs.~\ref{2xSR} and \ref{4xSR}, respectively. As compared to the state-of-the-art SISR and LF image SR methods, our method can produce images with more faithful details and less artifacts. Specifically, for 2$\times$SR, the images generated by our LF-DFnet are very close to the groundtruth images. Note that, the stairway in scene \textit{INRIA\_Sculpture} and the horizontal stripes in scene \textit{HCInew\_Origami} are faithfully recovered by our method without blurring or artifacts. For 4$\times$SR, state-of-the-art SISR methods EDSR and RCAN produce blurring results, and the perceptual-oriented SISR method ESRGAN generates images with fake textures. That is because, the SR problem becomes highly ill-posed for 4$\times$SR, and the spatial information in a single image is insufficient to reconstruct high-quality HR images. In contrast, our LF-DFnet can use complementary information among different views to recover missing details, and thus achieves better SR performance.

\begin{table}
\caption{Comparisons of the number of parameters (i.e., \#Params.), FLOPs, and reconstruction accuracy for $2\times$ and $4\times$ SR. Note that, FLOPs is calculated on an input LF with a size of $5\times5\times32\times32$. Here, we use PSNR and SSIM values averaged over 5 datasets \cite{EPFL,HCInew,HCIold,INRIA,STFgantry} to represent their reconstruction accuracy.}\label{tabEfficiency}
\centering
\renewcommand\arraystretch{1.1}
\begin{tabular}{|l|c|c|c|c|}
\hline
Method & Scale & \#Params. & FLOPs(G) &PSNR$/$SSIM \\[0.5pt]
\hline
EDSR \cite{EDSR}                & $2\times$ & 38.62M & 39.56$\times$25   & 36.04$/$0.9737 \\
RCAN \cite{RCAN}               & $2\times$ & 15.31M & 15.59$\times$25   & 36.11$/$0.9741 \\
resLF \cite{resLF}                  & $2\times$ & 6.35M   & 37.06                   & 36.57$/$0.9793 \\
LFSSR \cite{LFSSR}             & $2\times$ & 0.81M   & 25.70                   & 37.53$/$0.9835 \\
LF-InterNet \cite{LF-InterNet} & $2\times$ & 4.80M   & 47.46                   & 38.08$/$0.9847 \\
LF-DFnet (ours)                      & $2\times$ & 3.94M   & 57.22                   & 38.42$/$0.9854 \\
\hline
EDSR \cite{EDSR}                & $4\times$ & 38.89M & 40.66$\times$25 & 30.20$/$0.9121 \\
RCAN \cite{RCAN}              & $4\times$ & 15.36M & 15.65$\times$25  & 30.27$/$0.9133 \\
resLF \cite{resLF}                  & $4\times$ & 6.79M  & 39.70                   & 30.43$/$0.9223 \\
LFSSR \cite{LFSSR}            & $4\times$ & 1.61M   & 128.44                & 31.23$/$0.9345 \\
LF-InterNet \cite{LF-InterNet}& $4\times$ & 5.23M  & 50.10                   & 31.59$/$0.9387 \\
LF-DFnet (ours)                     & $4\times$ & 3.99M  & 57.31                   & 31.86$/$0.9415 \\
\hline
\end{tabular}
\end{table}

\subsubsection{Computational Efficiency}

We compare our LF-DFnet to several competitive methods (i.e., EDSR \cite{EDSR}, RCAN \cite{RCAN}, LFSSR \cite{LFSSR}, resLF \cite{resLF}, LF-InterNet \cite{LF-InterNet}) in terms of the number of parameters (i.e., \#Params) and FLOPs. As shown in Table~\ref{tabEfficiency}, our method achieves the highest PSNR and SSIM scores with a small number of parameters and FLOPs. Note that, the FLOPs of our method is significantly lower than EDSR and RCAN but slightly higher than resLF and LF-InterNet. That is because, our LF-DFnet uses more complicated feature extraction and reconstruction modules than resLF and LF-InterNet. These modules introduce a notable performance improvement at the cost of a reasonable increase of FLOPs. 
\begin{figure}
\centering
\includegraphics[width=8.8cm]{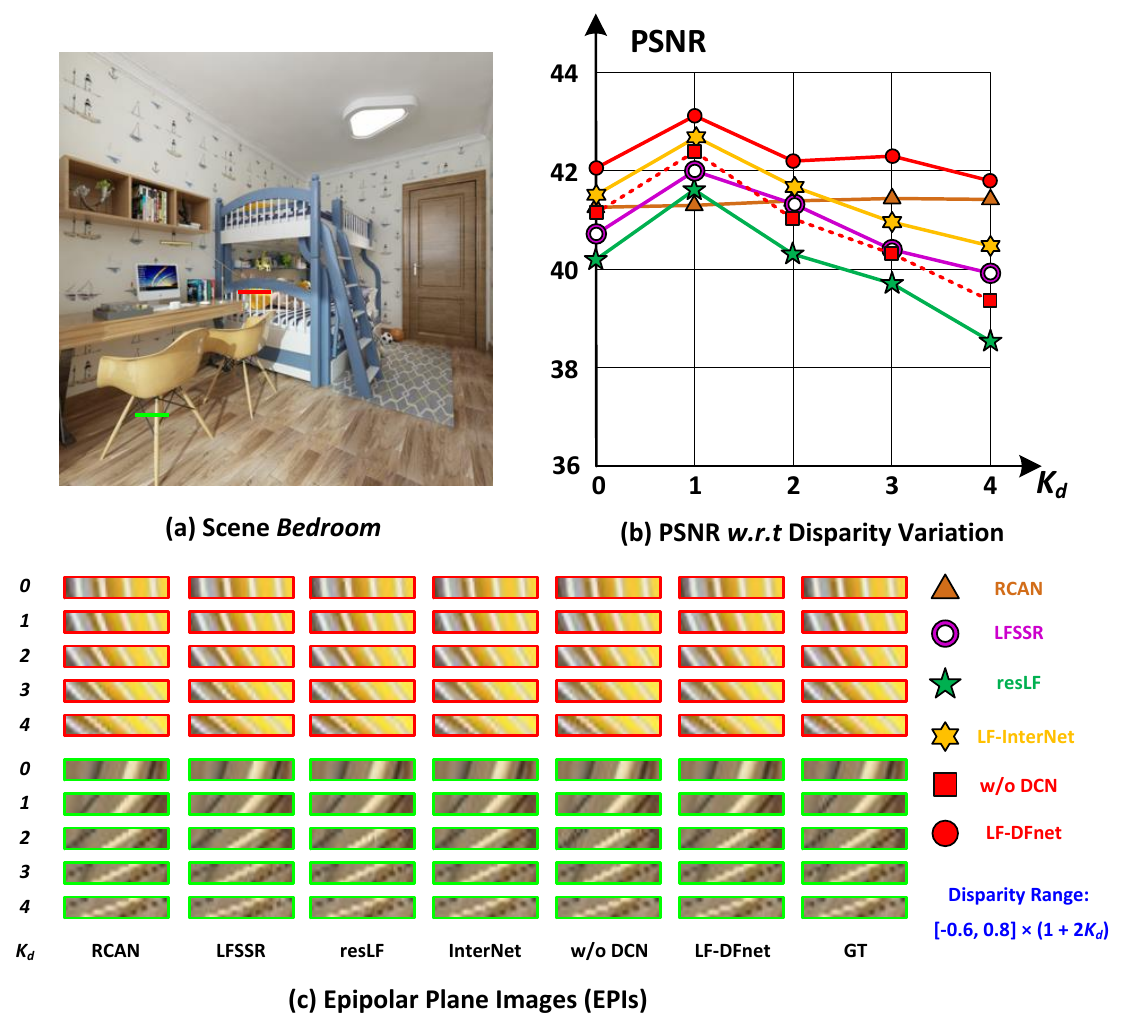}
\caption{Performance w.r.t. disparity variations. (a)~Scene \textit{Bedroom} of the NUDT dataset. (b)~PSNR values achieved by RCAN \cite{RCAN}, resLF \cite{resLF}, LFSSR \cite{LFSSR}, LF-InterNet \cite{LF-InterNet}, LF-DFnet, and LF-DFnet without deformable convolution (i.e., w/o DCN) under linearly increased disparities for 2$\times$SR. Our LF-DFnet achieves better SR performance than LFSSR and resLF, especially with large disparity variations (i.e., $K_{d}\geq2$), and the deformable convolution contributes to the performance improvements. (c)~Epipolar plane images (EPIs) of the corresponding strokes in (a). It can be observed that the misalignment becomes more severe as the baseline length (i.e., $K_d$) is increased.} \label{figPwrtB}
\end{figure}

\begin{figure*}
\centering
\includegraphics[width=18cm]{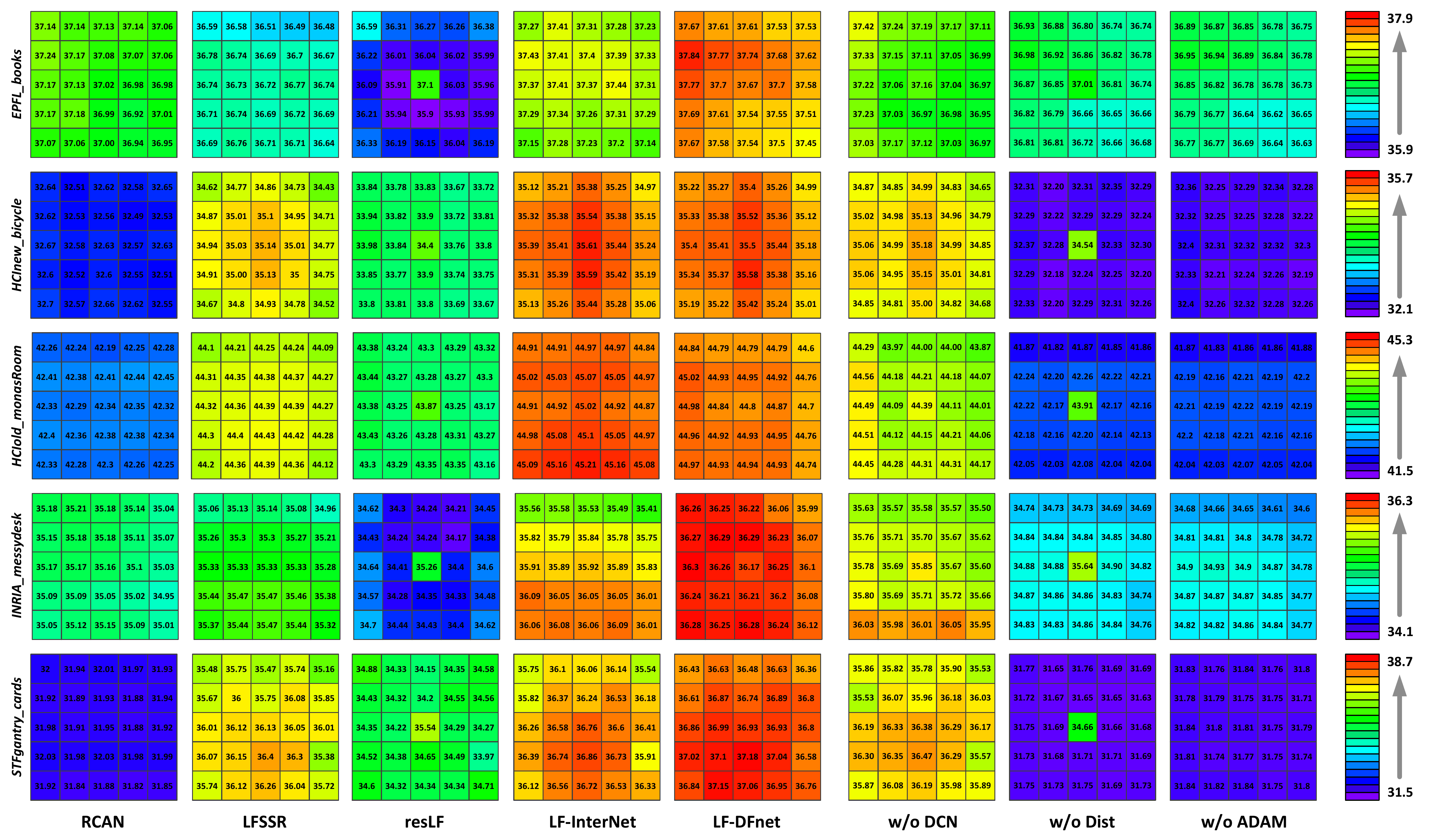}
\caption{A visualization of PSNR distribution among different perspectives on $5\times5$ LFs for 2$\times$SR. Here, we compare our LF-DFnet to 4 state-of-the-art SR methods (i.e., RCAN \cite{RCAN}, resLF \cite{resLF}, LFSSR \cite{LFSSR}, LF-InterNet \cite{LF-InterNet}) and three variants of our network (i.e., without using deformable convolution (\textit{w/o DCN}), without performing feature distribution (\textit{w/o Dist}), and without using ADAM (\textit{w/o ADAM})). Our LF-DFnet achieves high reconstruction quality with a balanced distribution among different perspectives.} \label{figPwrtP}
\end{figure*}

\subsubsection{Performance w.r.t. Disparity Variations}\label{secBaseline}
We selected scene \textit{Bedroom} (see Fig.~\ref{figPwrtB}(a)) from the NUDT dataset and rendered it with linearly increased baselines ($K_d = 0,1,\cdots,4$) to investigate the performance of LF image SR algorithms with respect to disparity variations. Note that, the disparities are proportional to the baseline length when the camera intrinsic parameters (e.g., focal length) are fixed. Following the HCInew dataset \cite{HCInew}, we calculated the disparity range (bottom-right in Fig.~\ref{figPwrtB}) using the corresponding groundtruth depth values. For performance evaluation, we first used RCAN \cite{RCAN} to evaluate inherent PSNR variation under different baseline lengths, and then compared our LF-DFnet to three state-of-the-art LF image SR methods (i.e., resLF \cite{resLF}, LFSSR \cite{LFSSR}, and LF-InterNet \cite{LF-InterNet}) and a variant of our network without using deformable convolution (i.e., w/o DCN). Details of this variant are introduced in Section~\ref{AblationADAM}. As shown in Fig.~\ref{figPwrtB}(b), RCAN achieves comparable SR performance under different baseline settings, which means that the inherent PSNR variation is low. The reconstruction accuracy of LFSSR and resLF drops significantly with increasing disparity variations. In contrast, our LF-DFnet is relatively insensitive to disparity variations and achieves comparable performance to RCAN when $K_{d}=4$. Note that, if deformable convolutions are replaced with regular 2D convolutions, our network (i.e., w/o DCN) suffers a notable performance degradation, especially under large baselines (e.g., $K_d>2$). That is because, large disparities can result in severe misalignment among LF images (see EPIs in Fig.~\ref{figPwrtB}(c)) and thus introduce difficulties in angular information exploitation. Since deformable convolution is used to perform feature alignment, our LF-DFnet is more robust to disparity variations, and thus achieves better performance on LF images with wide baselines.

\subsubsection{Performance w.r.t. Perspectives}\label{secPwrtP}
Since LF image SR methods aim at super-resolving all SAIs in an LF, we investigate the reconstruction accuracy of different methods with respect to different perspectives. For each dataset listed in Table~\ref{datasets}, we selected one scene from its test set and calculated the PSNR values on each SAI, as visualized in Fig.~\ref{figPwrtP}. We used $5\times5$ central SAIs to perform $2\times$SR, and compared our LF-DFnet to 4 state-of-the-art SR methods (i.e., RCAN \cite{RCAN}, LFSSR \cite{LFSSR}, resLF \cite{resLF}, LF-InterNet \cite{LF-InterNet}) and 3 variants of our network (i.e., without using deformable convolution (\textit{w/o DCN}), without performing feature distribution (\textit{w/o Dist}), and without using ADAM (\textit{w/o ADAM})). Details for these variants are introduced in Section~\ref{AblationADAM}. As shown in Fig.~\ref{figPwrtP}, resLF achieves high PSNR scores on center views but low PSNR scores on side views. That is because, resLF only uses a small part of views to super-resolve side views. The ignored angular information in these discarded views results in the imbalanced PSNR distribution. In contrast, both LF-InterNet and our LF-DFnet achieve improved reconstruction accuracy with a relatively balanced PSNR distribution. It is worth noting that, notable performance drop can be resulted by our LF-DFnet when DCN, ADAM, or feature distribution are removed or canceled. The above experiments clearly demonstrate the effectiveness of our ADAM and collect-and-distribute approach, and illustrate the high reconstruction quality of our LF-DFnet among different perspectives.

\begin{figure*}
\centering
\includegraphics[width=18cm]{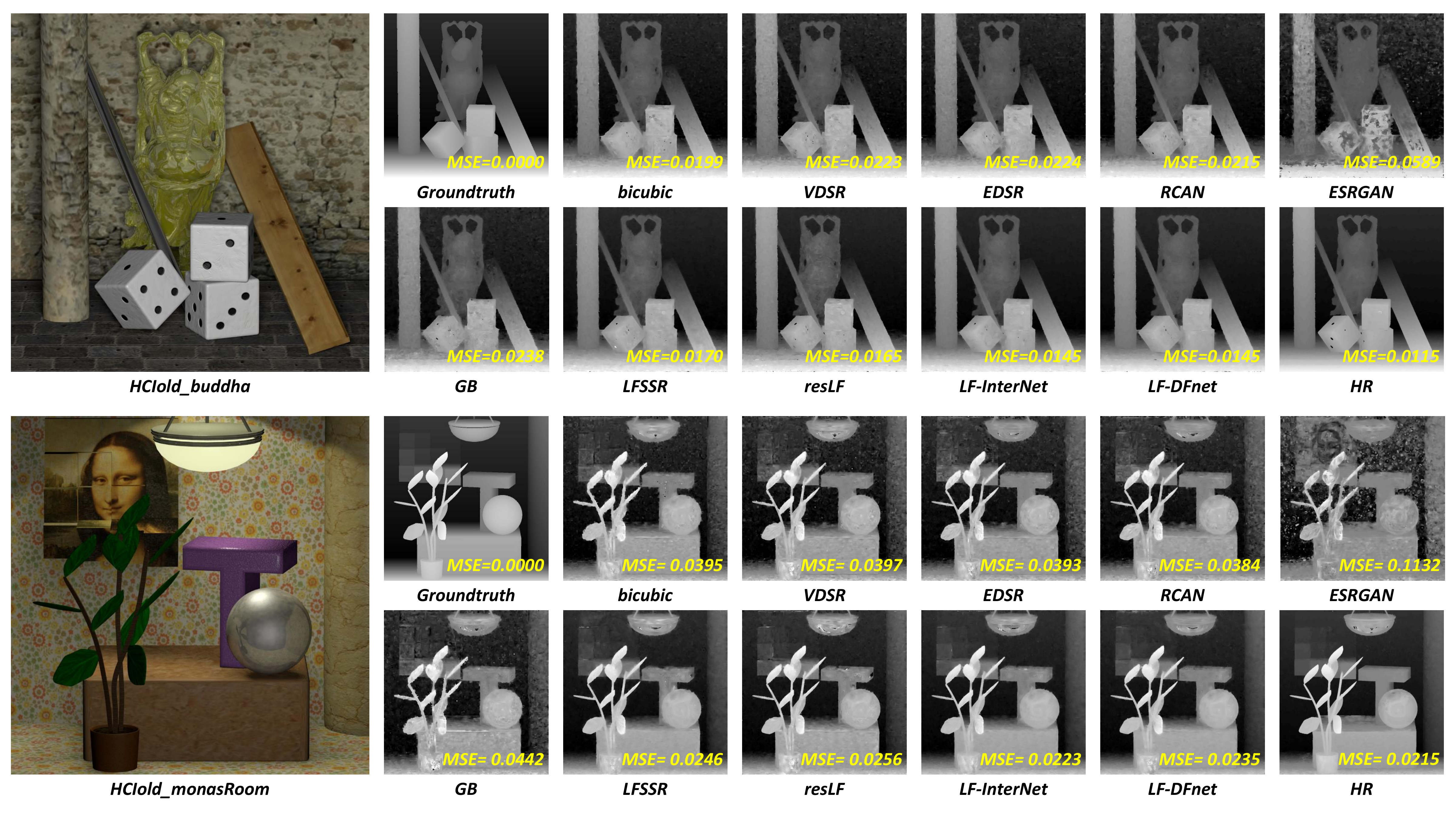}
\caption{Depth estimation results achieved by SPO \cite{SPO} method using 4$\times$SR LF images produced by different SR methods. Note that, the accuracy of depth estimation is improved by using the LF images produced by our LF-DFnet. That is, our LF-DFnet can generate angular-consistent HR LF images which are contributive to depth estimation.} \label{figDepth}
\end{figure*}

\begin{figure}
\centering
\includegraphics[width=8.8cm]{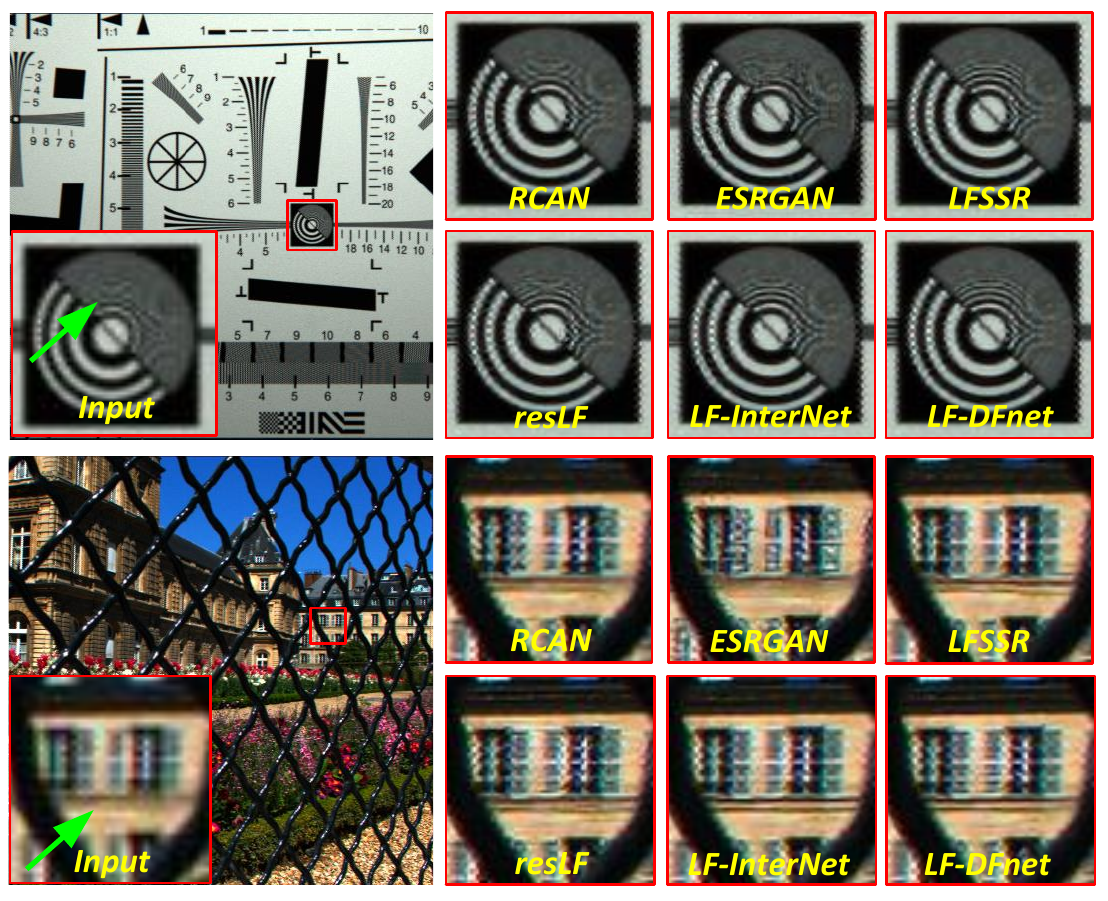}
\caption{Visual results achieved by different methods on real-world images.} \label{RealSR}
\end{figure}
\subsubsection{Performance on Real-World LF images}\label{secRealSR}
We compare our method to RCAN \cite{RCAN}, ESRGAN \cite{ESRGAN}, LFSSR \cite{LFSSR}, resLF \cite{resLF}, and LF-InterNet \cite{LF-InterNet} on real-world LF images by directly applying them to LFs in the EPFL dataset \cite{EPFL}. Since groundtruth HR images are unavailable in this case, we compare the visual performance of different methods. As shown in Fig.~\ref{RealSR}, our LF-DFnet recovers finer details than RCAN, and produces less artifacts than ESRGAN. It demonstrates that our method can be applied to LF cameras to generate high-quality HR images.

\begin{table*}
\caption{
PSNR$/$SSIM values achieved by LF-DFnet and its variants for $2\times$SR.}\label{tabAblation}
\centering
\scriptsize
\renewcommand\arraystretch{1.2}
\begin{tabular}{|l|c|ccccc|c|}
\hline
\multirow{2}*{Model}&\multirow{2}*{\#Params.}&  \multicolumn{5}{c|}{Dataset}& \multirow{2}*{Average} \\
\cline{3-7}
  &  & EPFL \cite{EPFL} &HCInew \cite{HCInew} & HCIold \cite{HCIold} & INRIA \cite{INRIA} & STFgantry \cite{STFgantry} &\\
\hline
\textit{LF-DFnet w/o DCN}                      & 3.77M & 34.04$/$0.9755 & 36.94$/$0.9761 & 43.63$/$0.9936 & 35.93$/$0.9837 & 38.73$/$0.9905 & 37.85$/$0.9839 (-0.57$/$-0.0015)\\
\textit{LF-DFnet w/o ADAM}                   & 4.21M & 32.81$/$0.9614 & 34.73$/$0.9585 & 40.94$/$0.9872 & 34.70$/$0.9749 & 36.40$/$0.9819 & 35.92$/$0.9728 (-2.50$/$-0.0126)\\
\textit{LF-DFnet w/o Dist}                        & 4.07M & 32.87$/$0.9616 & 34.89$/$0.9614 & 41.15$/$0.9889 & 34.81$/$0.9765 & 36.57$/$0.9831 & 36.02$/$0.9743 (-1.85$/$-0.0111)\\
\textit{LF-DFnet w/o ASPPinFEM}           & 3.96M & 34.32$/$0.9762 & 37.21$/$0.9774 & 43.89$/$0.9941 & 36.24$/$0.9837 & 39.37$/$0.9928 & 38.21$/$0.9848 (-0.21$/$-0.0006)\\
\textit{LF-DFnet w/o ASPPinOFB}           & 4.01M & 34.36$/$0.9764 & 37.33$/$0.9781 & 43.94$/$0.9941 & 36.20$/$0.9838 & 38.95$/$0.9923 & 38.16$/$0.9849 (-0.26$/$-0.0005)\\
\textit{\textbf{LF-DFnet}}                         & 3.94M & \textbf{34.44$/$0.9766} & \textbf{37.44$/$0.9786} & \textbf{44.23$/$0.9943} & \textbf{36.36$/$0.9841} & \textbf{39.61$/$0.9935} & \textbf{38.42$/$0.9854 ( 0.00$/$ 0.0000)}\\
\hline
\end{tabular}
\end{table*}
\subsubsection{Benefits to Depth Estimation}
Since high-resolution and angular-consistent LF images are beneficial to depth estimation, we evaluate the reconstruction quality and angular consistency of different SR methods by using their generated LFs to perform depth estimation. Specifically, we used the methods presented in Fig.~\ref{4xSR} to perform $4\times$SR on the HCIold dataset \cite{HCIold}, and applied SPO method \cite{SPO} to the super-resolved LF images to perform depth estimation. Note that, the original HR LFs and LFs generated by bicubic interpolation method were used to produce upper bound results and baseline results, respectively. The mean square error (MSE) metric was used to measure the distance between estimated disparities and groundtruth disparities (normalized to $[0,1]$). It can be observed in Fig.~\ref{figDepth} that, although SISR methods \cite{VDSR,EDSR,RCAN,ESRGAN} achieve superior quantitative$/$visual performance over the bicubic interpolation method (as shown in Table~\ref{tabComparison} and Fig.~\ref{4xSR}), they achieve comparable or even worse results in depth estimation. That is because, SISR methods
super-resolve each SAI separately without using any angular information. Consequently, these SISR methods cannot ensure the angular consistency in their generated LFs, which is significantly important to depth estimation. In contrast, several LF image SR methods (i.e., LFSSR, resLF, LF-InterNet, LF-DFnet) can improve the depth estimation performance by generating high-resolution and angular consistent LF images. Note that, the depth estimation results using the output of LF-InterNet and LF-DFnet are very close to the results using the original HR LFs, which clearly demonstrates the high spatial reconstruction quality and angular consistency achieved by these two methods.

\subsection{Ablation Study}\label{secAblation}
In this subsection, we compare our LF-DFnet with several variants to investigate the potential benefits introduced by our network modules and design choices.
\subsubsection{Angular Deformable Alignment Module (ADAM)}\label{AblationADAM}
 As the core component of our LF-DFnet, ADAM can perform  bidirectional feature alignment between the center view and each side view using deformable convolutions. Here, we validate the effectiveness of ADAM by introducing the following three variants:
 \begin{itemize}
 \item \textbf{LF-DFnet w/o DCN}: We replaced the deformable convolution with regular $3\times3$ convolution in both feature collection and feature distribution stages. Note that, both network depth and feature width of this variant are the same as the original network. Since the offset learning branch was removed with deformable convolution, this variant has a 0.17M parameter reduction as compared to \textit{LF-DFnet}.
 \item \textbf{LF-DFnet w/o ADAM}: We removed all ADAMs in this variant to investigate their contributions. Specifically, we removed both feature collection and feature distribution, and used a residual block (i.e., two $3\times3$ convolutions and a Leaky ReLU layer) to process each view separately. To achieve fair comparison (i.e., comparable network depth and model size), we increased the number of filters of all convolution layers to make its model size slightly larger than \textit{LF-DFnet}.
 \item \textbf{LF-DFnet w/o Dist}: To investigate the benefit introduced by the bidirectional feature interaction mechanism, we removed feature distribution and only performed feature collection in this variant. Specifically, we used deformable convolution to align side-view features to center view and performed feature fusion as in \textit{LF-DFnet}. However, we did not distribute the incorporated features to side views but followed \textit{LF-DFnet w/o ADAM} to process each side-view feature separately. Similar to \textit{LF-DFnet w/o ADAM}, we adjusted the number of filters to make its model size not smaller than \textit{LF-DFnet}.
  \end{itemize}

  Table~\ref{tabAblation} shows the comparative results achieved by \textit{LF-DFnet} and its variants. It can be observed in the table that the average PSNR value of \textit{LF-DFnet w/o DCN} suffers a decrease of 0.57 dB as compared to \textit{LF-DFnet}, which demonstrates the effectiveness of deformable convolution in \textit{LF-DFnet}. It is worth noting that, the performance degradation is more significant for the dataset with wide baselines (see \textit{w/o DCN} in Fig.~\ref{figPwrtB} and scores on the STFgantry dataset in Table~\ref{tabAblation}). That is because, wide baselines can cause large disparity variations and thus result in severe misalignments among different SAIs. Consequently, the contributive angular information cannot be effectively incorporated without using deformable convolution for feature collection and distribution.

  When all ADAMs are removed from \textit{LF-DFnet}, the network (i.e., \textit{LF-DFnet w/o ADAM}) is identical to an SISR model which only uses spatial information within single views to separately super-resolve each SAI. As shown in Table~\ref{tabAblation}, \textit{LF-DFnet w/o ADAM} achieves 35.92 dB in average PSNR, which is significantly lower than \textit{LF-DFnet} (38.42 dB) but marginally higher than VDSR \cite{VDSR} (35.49 dB with a 0.66M model). This clearly demonstrates the importance of angular information in LF image SR.

  Finally, when feature distribution is canceled, the angular information can be only propagated from side views to center view by \textit{LF-DFnet w/o Dist}, which results in a decrease of 1.85 dB in average PSNR as compared to \textit{LF-DFnet}. It is worth noting that, although angular information is used to super-resolve the center view, as shown in Fig.~\ref{figPwrtP}, the PSNR values of \textit{LF-DFnet w/o Dist} on center views are still much lower than those of \textit{LF-DFnet}. That is because, the parameters of the reconstruction module are shared among different perspectives. The unidirectional feature propagation (only from side view to center view) makes the center-view feature significantly vary from the side-view features. This asymmetric feature distribution hinders the reconstruction module to achieve high reconstruction accuracy on both center view and side views.

\begin{table*}
\caption{
PSNR$/$SSIM values achieved by LF-DFnet with different number of ADAMs for $2\times$SR.}\label{tabAblationADAMs}
\centering
\scriptsize
\renewcommand\arraystretch{1.2}
\begin{tabular}{|l|c|ccccc|c|}
\hline
\multirow{2}*{Model}&\multirow{2}*{\#Params.}&  \multicolumn{5}{c|}{Dataset}& \multirow{2}*{Average} \\
\cline{3-7}
  &  & EPFL \cite{EPFL} &HCInew \cite{HCInew} & HCIold \cite{HCIold} & INRIA \cite{INRIA} & STFgantry \cite{STFgantry} &\\
\hline
\textit{LF-DFnet with 1ADAM}                      & 2.52M & 33.69$/$0.9734 & 36.55$/$0.9760 & 43.82$/$0.9938 & 35.96$/$0.9828 & 39.12$/$0.9921 & 37.83$/$0.9835 (-0.59$/$-0.0019)\\
\textit{LF-DFnet with 2ADAMs}                    & 3.23M & 34.40$/$0.9765 & 37.38$/$0.9784 & 44.06$/$0.9941 & 36.33$/$0.9841 & 39.40$/$0.9932 & 38.31$/$0.9853 (-0.11$/$-0.0001)\\
\textit{\textbf{LF-DFnet} (3 ADAMs)}           & 3.94M & 34.44$/$0.9766 & 37.44$/$0.9786 & 44.23$/$0.9943 & 36.36$/$0.9841 & 39.61$/$0.9935 & 38.42$/$0.9854 ( 0.00$/$ 0.0000)\\
\textit{LF-DFnet with 4ADAMs}                    & 4.65M & 34.47$/$0.9767 & 37.51$/$0.9787 & 44.34$/$0.9944 & 36.35$/$0.9841 & 39.76$/$0.9937 & 38.49$/$0.9855 ( 0.07$/$ 0.0001)\\
\hline
\end{tabular}
\end{table*}
\subsubsection{Residual ASPP Module}
Residual ASPP module is used in our LF-DFnet for both feature extraction and offset learning. To demonstrate its effectiveness, we introduced two variants (i.e., \textit{LF-DFnet w/o ASPPinFEM} and \textit{LF-DFnet w/o ASPPinOFB}) by replacing the residual ASPP blocks with residual blocks in the feature extraction module and the offset learning branch, respectively. As shown in Table~\ref{tabAblation}, \textit{LF-DFnet w/o ASPPinFEM} suffers a 0.21 dB decrease in average PSNR as compared to \textit{LF-DFnet}. That is because, residual ASPP module can extract hierarchical features from input images, which are beneficial to LF image SR. Similarly, a 0.26 dB PSNR decrease is introduced when ASPP module is removed from the offset learning branch. That is because, the ASPP module can achieve accurate offset learning through multi-scale feature representation and the enlargement of receptive fields.

\subsubsection{Number of ADAMs}\label{secAblationADAMs}

We investigate the SR performance with respect to the number of ADAMs in our network. It can be observed in Table~\ref{tabAblationADAMs} that the reconstruction accuracy consistently improves as the number of ADAMs increases. However, the improvements tend to be saturated when the number of ADAMs is increased from 3 to 4. Since the number of parameters grows linearly with respect to the number of ADAMs, we decided to use 3 ADAMs (i.e., $K=3$) in our LF-DFnet to achieve a good tradeoff between reconstruction accuracy and computational efficiency.

\section{Conclusion and Discussion}\label{secConclusion}
In this paper, we propose an LF-DFnet to achieve LF image SR. Different from existing LF image SR methods, we explicitly handle the disparity problem by performing feature alignment using our designed angular deformable alignment module (ADAM).  Moreover, we developed the first baseline-adjustable LF dataset in literature to evaluate the performance of LF image SR methods with respect to disparity variations. Extensive experiments on both public and our self-developed datasets have demonstrated the effectiveness of the proposed method. Our LF-DFnet achieves state-of-the-art SR performance with a small computational cost. The reconstruction accuracy achieved by our LF-DFnet is evenly distributed among angular views and is more robust to disparity variations.

Moreover, as demonstrated in the experiments, our LF-DFnet works well on real-world LF images, and can be used to generate high-quality (i.e., high-resolution and angular-consistent) LFs to benefit downstream tasks (e.g., LF depth estimation). Since high-resolution LF images are needed in many LF tasks (e.g., post-capture refocusing, depth sensing), the experimental results in this paper clearly demonstrates the promising potential applications of our LF-DFnet. In the future, we will try to combine our LF-DFnet with other task-specific networks to further boost the performance gain introduced by high-resolution LF images.

It is worth noting that, although our method achieves promising reconstruction accuracy in terms of PSNR and SSIM values, the visual superiority of our LF-DFnet is very minor and far from satisfactory. That is because, only $L_1$ loss was used to train our network, and the quality of input LR images are very low, especially for those captured by Lytro cameras.  In the future, we will exploit the generative adversarial network (GAN) paradigm to improve the visual quality of reconstructed images and achieve LF image SR with large scaling factors (e.g., 16$\times$ SR). We believe that GAN-based LF image SR networks can achieve a better perception-distortion tradeoff by using the additional angular information provided by LF images, and will take a further step toward consumer applications.
\bibliographystyle{IEEEtran}
\bibliography{LF-DFnet}

\begin{IEEEbiography}[{\includegraphics[width=1in,height=1.25in,clip,keepaspectratio]{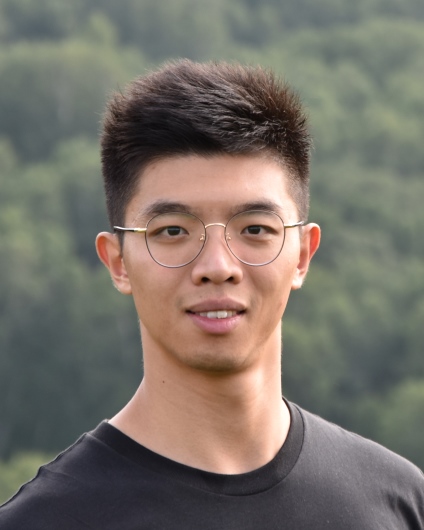}}]{Yingqian Wang} received the B.E. degree in electrical engineering from Shandong University (SDU), Jinan, China, in 2016, and the M.E. degree in information and communication engineering from National University of Defense Technology (NUDT), Changsha, China, in 2018. He is currently pursuing the Ph.D. degree with the College of Electronic Science and Technology, NUDT. His research interests focus on low-level vision, particularly on light field imaging and image super-resolution.
\end{IEEEbiography}

\begin{IEEEbiography}[{\includegraphics[width=1in,height=1.25in,clip,keepaspectratio]{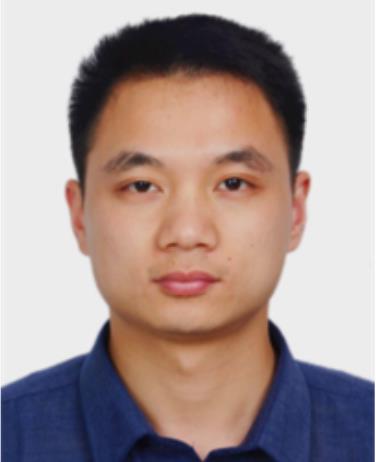}}]{Jungang Yang} received the B.E. and Ph.D. degrees from National University of Defense Technology (NUDT), in 2007 and 2013 respectively. He was a visiting Ph.D. student with the University of Edinburgh, Edinburgh from 2011 to 2012. He is currently an associate professor with the College of Electronic Science, NUDT. His research  interests include computational  imaging, image processing, compressive sensing and sparse representation. Dr. Yang received the New Scholar Award of Chinese Ministry of Education in 2012, the Youth Innovation Award and the Youth Outstanding Talent of NUDT in 2016.
\end{IEEEbiography}

\begin{IEEEbiography}[{\includegraphics[width=1in,height=1.25in,clip,keepaspectratio]{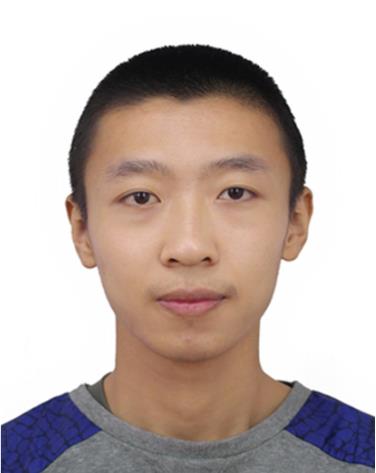}}]{Longguang Wang} received the B.E. degree in electrical engineering from Shandong University (SDU), Jinan, China, in 2015, and the M.E. degree in information and communication engineering from National University of Defense Technology (NUDT), Changsha, China, in 2017. He is currently pursuing the Ph.D. degree with the College of Electronic Science and Technology, NUDT. His research interests include low-level vision and deep learning.
\end{IEEEbiography}

\begin{IEEEbiography}[{\includegraphics[width=1in,height=1.25in,clip,keepaspectratio]{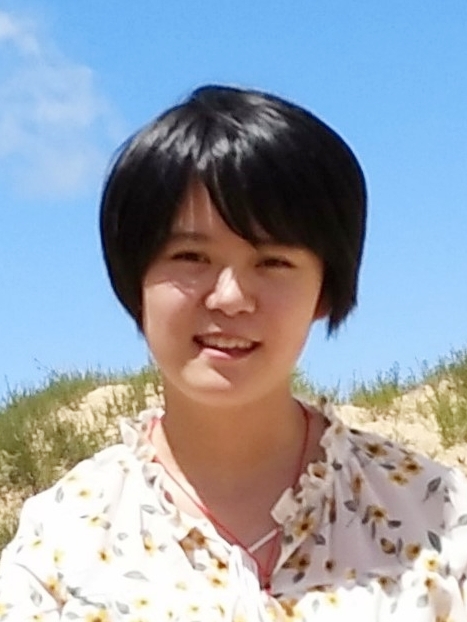}}]{Xinyi Ying} is currently pursuing the M.E. degree with the College of Electronic Science and Technology, National University of Defense Technology (NUDT). Her research interests focus on low-level vision, particularly on image and video super-resolution.
\end{IEEEbiography}

\begin{IEEEbiography}[{\includegraphics[width=1in,height=1.25in,clip,keepaspectratio]{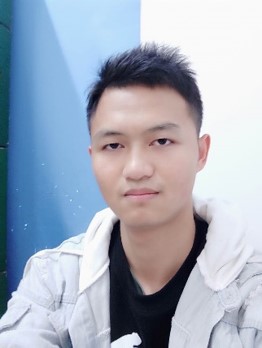}}]{Tianhao Wu} received the B.E. degree  in electronic engineering from National University of Defense Technology (NUDT), Changsha, China, in 2020. He is currently pursuing the M.E. degree with the College of Electronic Science and Technology, NUDT. His research interests include light field imaging and camera calibration.
\end{IEEEbiography}

\begin{IEEEbiography}[{\includegraphics[width=1in,height=1.25in,clip,keepaspectratio]{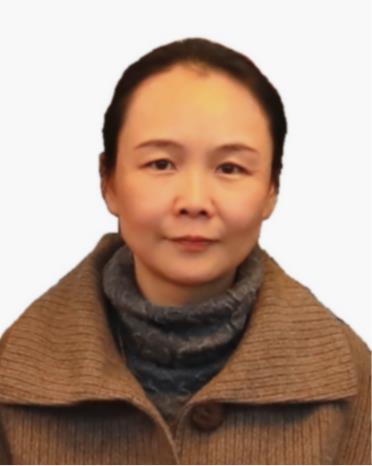}}]{Wei An} received the Ph.D. degree from the National University of Defense Technology (NUDT), Changsha, China, in 1999. She was a Senior Visiting Scholar with the University of Southampton, Southampton, U.K., in 2016. She is currently a Professor with the College of Electronic Science and Technology, NUDT. She has authored or co-authored over 100 journal and conference publications. Her current research interests include signal processing and image processing.
\end{IEEEbiography}

\begin{IEEEbiography}[{\includegraphics[width=1in,height=1.25in,clip,keepaspectratio]{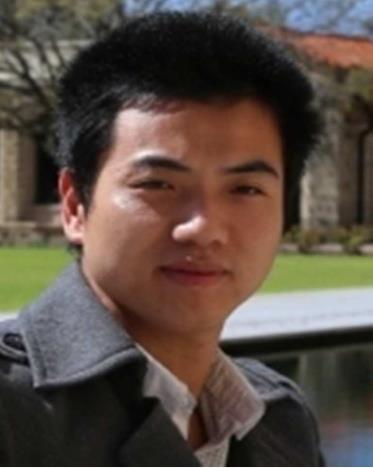}}]{Yulan Guo} is currently an associate professor. He received the B.E. and Ph.D. degrees from National University of Defense Technology (NUDT) in 2008 and 2015, respectively. He was a visiting Ph.D. student with the University of Western Australia from 2011 to 2014. He worked as a postdoctorial research fellow with the Institute of Computing Technology, Chinese Academy of Sciences from 2016 to 2018. He has authored over 90 articles in journals and conferences, such as the IEEE TPAMI and IJCV. His current research interests focus on 3D vision, particularly on 3D feature learning, 3D modeling, 3D object recognition, and scene understanding. Dr. Guo received the CAAI Outstanding Doctoral Dissertation Award in 2016, the CAAI Wu-Wenjun Outstanding AI Youth Award in 2019. He served/will serve as an associate editor for IET Computer Vision and IET Image Processing, a guest editor for IEEE TPAMI, an area chair for CVPR 2021 and ICPR 2020, an organizer for a tutorial in CVPR 2016 and a workshopin CVPR 2019. Dr. Guo is a senior member of IEEE.
\end{IEEEbiography}

\end{document}